\documentclass[journal,comsoc]{IEEEtran}
\pdfoutput=1

\usepackage[T1]{fontenc}
\usepackage{gensymb}
\usepackage{todonotes}
\usepackage{booktabs}
\usepackage{multirow}
\usepackage{amsmath}
\usepackage{afterpage}
\usepackage[noadjust]{cite}

\makeatletter
\if@twocolumn%
\fi
\makeatother

\usepackage{amsmath}
\usepackage{algorithm}
\usepackage{algpseudocode}

\makeatletter
\def\BState{\State\hskip-\ALG@thistlm}
\makeatother

\usepackage{bm}

\usepackage{url}
\usepackage{bbm}
\usepackage{graphicx}

\begin{document}

\title{Spectrum and Infrastructure Sharing in Millimeter Wave Cellular Networks: An Economic Perspective}

\author{Fraida~Fund, Shahram~Shahsavari, Shivendra~S.~Panwar, Elza~Erkip, Sundeep~Rangan\thanks{All of the authors are with the Department of Electrical and Computer Engineering, NYU Tandon School of Engineering.}}%
\maketitle

\begin{abstract}
The licensing model for millimeter wave bands has been the subject of considerable
debate, with some industry players advocating for unlicensed use and others
for traditional geographic area exclusive use licenses. Meanwhile, 
the massive bandwidth, highly directional antennas, high penetration 
loss and susceptibility to shadowing in these bands suggest certain
advantages to spectrum and infrastructure sharing. However, even 
when sharing is technically beneficial (as recent research in this area 
suggests that it is), it may not be profitable. In this paper, 
both the technical and economic implications of resource 
sharing in millimeter wave networks are studied. Millimeter wave 
service is considered in the economic framework of a network good,
where consumers' utility depends on the size of the network, 
and the strategic decisions of consumers and service providers
are connected to detailed network simulations. The results suggest 
that ``open'' deployments of neutral small cells that serve subscribers
of any service provider encourage market entry by making it easier for 
networks to reach critical mass, more than ``open'' (unlicensed) 
spectrum would.  The conditions under which 
competitive service providers would prefer to share resources 
or not are also described.
\end{abstract}

\IEEEpeerreviewmaketitle

\section{Introduction}

The millimeter wave (mmWave) bands represent one of
the largest unlicensed bandwidths ever allocated,
presenting a tremendous opportunity for both
technical and policy innovation.
The appropriate licensing model for this band
remains the subject of considerable debate.
Replies to an FCC notice of inquiry~\cite{noi-fcc}
requesting comments on usage of bands greater than 24 GHz in the United States
reveal disagreement on how to best utilize this spectrum,
with economic considerations playing a significant role.
Major industry players argued in favor of exclusive use licensing on a geographic
service area basis, primarily on the grounds that this offers sufficient
certainty to motivate major capital
investment. Several of these explicitly asked the FCC
to reject licensing mechanisms that require spectrum sharing
on some 
bands~\cite{comment-tmobile-fcc,comment-verizon-fcc,comment-straightpath-fcc,
comment-qualcomm-fcc,comment-ctia-fcc,comment-nokia-fcc,comment-att-fcc}.
Others argued that unlicensed use maximizes efficient spectrum use, and
encourages innovation and competition by
lowering barriers to entry~\cite{comment-ncta-fcc,comment-google-fcc}.
A recent notice of proposed rulemaking~\cite{nprm-fcc} for these bands
involves 3,850 MHz of spectrum, but does not move on an additional 12,500 MHz of
potentially useful spectrum in bands above 24 GHz.

Beyond these business concerns, technical properties
of mmWave bands favor spectrum and infrastructure (base station) sharing.
While cellular frequencies have traditionally been allocated
with geographic area exclusive use licenses, the
physical characteristics of mmWave signals
suggest that exclusive use licenses would be sub-optimal in
these bands. Specifically, in the mmWave space, the massive bandwidth
and spatial degrees of freedom are unlikely to be fully used by
any one cellular operator.  The use of high-dimensional
antenna arrays implies that spectrum can be shared,
not just in time, but also in space.
Furthermore, mmWave signals observe
high penetration loss through brick and
glass~\cite{Rappaport2014-mmwbook}, and
are highly susceptible to shadowing.
This implies that many more base stations are
likely to be needed for wide area coverage, 
significantly increasing the cost of deployment, thus
motivating infrastructure sharing.

However, technological justification for resouce sharing
does not always translate to economic benefits, for service
providers or for consumers.
Network service providers are mainly concerned with increasing
profit, which is a function of demand, price, and cost.
Even when resource sharing improves consumers' quality of service,
it may have a negative effect on the service provider's profits
if it shifts demand to a competing service provider, 
or if it changes the market dynamics in a way 
that forces down the price.
Similarly, consumers prefer a higher quality of service,
but they are also concerned with service availability
and price, which could potentially be negatively affected by resource sharing.
To gain a fuller understanding of the benefits of resource
sharing in mmWave networks,
we need to identify the specific impact on
quality of service, and then understand how this
affects the demand, price, and cost of service.

\subsection{Contributions}

The goal of this work is to model the strategic decisions of
wireless service providers in building out mmWave networks
with or without sharing of spectrum and/or infrastructure.
We apply economic models of network
goods~\cite{economides1996economics} - products whose
value to consumers depends on the number of units sold - to mmWave cellular
networks, where the value of the network to the consumer
depends on the size of the network (in terms of base station
and spectrum resources), and the investment
of the service provider in base station and spectrum resources
depends on its expected market share.
We quantify the positive and negative effects
associated with increasing network size, 
i.e., how a subscriber's data rate changes
as the mmWave service provider increases its spectrum holdings,
base station deployments, and market share.
Using the concept of \emph{critical mass}~\cite{economides1995critical}, 
we investigate the growth of demand for a mmWave network service
under three circumstances: increasing a network from 
zero size by deploying base stations and licensing spectrum,
licensing spectrum but utilizing an existing deployment
of ``open'' small cells,
and deploying base stations but using unlicensed spectrum.
We further model the resource sharing 
decision of competing mmWave service providers as a 
\emph{compatibility problem}~\cite{katz1985network,economides1998equilibrium}.

The contributions of this work are primarily to connect
the performance of mmWave networks to economic models of
demand for these network services, as follows:
\begin{itemize}
\item We quantify the positive and negative effects
of subscribing to a large service provider -
one with greater base station density, bandwidth, and 
number of subscribers - as well as interactions between these.  
We find a strong positive
effect with increasing base station density
in a low-SNR regime, but only a weak
positive effect with increasing bandwidth.
In high-SNR regimes, there is only a weak positive effect with increasing
base station density, and this is partially mitigated by a very slight
negative interference effect.
Moderate-SNR subscribers gain the most from increasing bandwidth and base station density.
\item Given these network effects, 
we consider the effect of ``open'' resources on new service offerings,
where the primary concern of the provider is to establish 
a stable presence in the market.
We find that the slow initial growth of demand at small network sizes
makes it difficult for a new provider to reach critical mass.
With an existing deployment of ``open'' small cells, there
is robust demand even at small network sizes, which encourages
growth. Open spectrum (i.e., unlicensed) does not have as
encouraging an effect on market entry.
\item We separately consider resource sharing between mmWave service providers
who are already established in the market, and are mainly concerned with profit. 
We apply the economic model of compatibility
of network goods to 
resource sharing in established mmWave cellular networks.
We describe a duopoly game involving two vertically differentiated
mmWave service providers with and without resource sharing,
and quantify the service provider profits and market coverage in each
case. We find that providers may prefer to share spectrum and 
base stations when the market is highly segmented, share base stations 
when the spectrum is unlicensed, and share spectrum when base
stations are all ``open''. Otherwise the high-end service provider
will prefer not to share resources.
\end{itemize}

\subsection{Related Work}

The idea of resource sharing is, of course, not new;
a great deal of research effort has been devoted
to quantifying the benefits of base station and spectrum
sharing in cellular networks.
In \cite{traditional}, the
authors consider several sharing options for LTE networks,
and conclude that an arrangment similar to a traditional roaming agreement
offers the best performance with the least complexity for inter-operator sharing.
The authors in \cite{shiv-sharing} assess the benefit of sharing both infrastructure and
spectrum in the context of a proposed merger between
two major cellular operators in the United States, using real base station deployment data
to support their claims.
In \cite{dublin} the authors investigate the trade-offs between
infrastructure sharing (which improves coverage and has a small positive effect on data rate)
and spectrum sharing (which has a positive effect on data rate but reduces coverage probability), and find that combining both kinds of sharing
offers the best data rate while partially mitigating the
reduced coverage of spectrum sharing.
All of these, as well as others \cite{3GPP-sharing-implementation,orthogonal-sharing},
conclude that under some conditions,
resource sharing increases the capacity of traditional cellular networks, but some find that
spectrum sharing without coordination in traditional cellular networks
creates interference and degrades performance relative to exclusive use of spectrum by one operator.

Given the unique propagation characteristics of mmWave networks,
there has been renewed interest in resource sharing in these bands.
mmWave networks will require a denser deployment of base stations
than conventional frequencies, increasing the appeal of base station sharing.
Industry perspectives on 5G cellular networks~\cite{industry-5g}
suggest a favorable view of neutral small cells owned by a third
party and shared by multiple operators.
With respect to spectrum sharing, as opposed to conventional cellular frequencies,
the inter-cell interference in mmWave bands can be controlled by
directional transmissions \cite{andrews-sharing,heath-mm,european-sharing},
potentially allowing much greater spectrum reuse.
In \cite{andrews-sharing}, it is shown that with sufficient
beam directionality in the transmission pattern,
the inter-cell interference is low enough to favor
resource sharing
even without inter-operator coordination.
These conclusions are supported by \cite{matia,matiamag} for
different channel models.
However, in \cite{european-sharing}, the authors show
that inter-operator coordination is important to users
with poor data rates, especially in dense deployments.
Similarly, \cite{coordination-sharing} proposes a spectrum sharing
scheme with inter-cell coordination to avoid
inter-cell interference and increase sharing gain.

With the exception of \cite{dublin}, all
of these have considered resource sharing only in the case
of symmetric service providers (those with equivalent spectrum
and base station resources, and equal market share).
On conventional cellular frequencies,
\cite{dublin} shows that resource sharing benefits
a large service provider more than
a small service provider, because less interference
is imposed by the small service provider.
However, in mmWave frequencies, where the effect of
intercell interference is much smaller, it is not clear
whether this result remains relevant.

Furthermore, even when full sharing is strictly beneficial
from a technical perspective, competitive dynamics between
service providers may discourage sharing unless there are external
incentives. With full resource sharing,
subscribers of all service providers have exactly the same quality of service,
making it difficult for service providers to distinguish
themselves in the market and gain market share.
None of~\cite{andrews-sharing,dublin,matia,european-sharing}
consider this effect.
In \cite{andrews-sharing}, the authors claim that with resource sharing,
a network operator requires less bandwidth (and therefore, lower
spectrum licensing costs) to serve its subscribers with a given median rate.
However, this assumes that demand for network services is fixed,
then it actually varies according to the quality of service,
and it also ignores competition between service providers.
The early work on mmWave resource sharing also does
not address the case of asymmetric service providers,
where the large service provider contributes more resources
to the partnership, but then offers
its subscribers the same quality of service as the small service provider.

Some of the literature on cellular networks addresses 
economic or regulatory aspects of resource sharing. 
For example,~\cite{dublin2} models the 
tradeoff associated with competition regulation and 
resource sharing in the context of the planned evolution 
of cellular networks. A coalition game described in~\cite{coalitionsharing}
suggests that resource sharing and cooperation
can sometimes improve individual cellular service providers' payoff.
A ten-year case study on cellular service providers in 
Sweden~\cite{Markendahl2012opetition}
lists incentives, obstacles, 
and key drivers for cooperation, and a similar investigation 
of cellular infrastructure sharing in emerging markets
is described in~\cite{Meddour:2011:RIS:1975024.1975472}.
However, none of these address resource sharing in mmWave networks which,
as mentioned above, are fundamentally different from 
previous cellular networks in ways that 
can affect the decision to share resources or not.
An early economic perspective on mmWave networks (although not on 
resource sharing) in~\cite{Nikolikj:2015:SBP:2822170.2822198}
suggests that the limited coverage range of mmWave-based 
5G systems is a key challenge for its cost efficiency.
Resource sharing could potentially be a way to address this challenge, 
but the economic implications of resource sharing in mmWave
networks have not been studied yet.

\subsection{Paper Organization}

The rest of this paper is organized as follows.
We begin with a brief introduction to the economic
framework used in this paper, in Section~\ref{sec:economic}.
In Section~\ref{sec:technical}, we describe the system model
and simulation results showing the benefit of resource sharing
in mmWave networks with respect to fifth percentile rate. Section~\ref{sec:network}
describes mmWave service as a network good (in the economic sense),
and uses simulation results to quantify the network externalities
associated with increasing network size.
We build on results from Section~\ref{sec:technical}
and Section~\ref{sec:network} in Section~\ref{sec:demand}
to show how demand for mmWave network
services evolves as a service provider increases its network size,
and we compare the likelihood of market entry with and without
``open'' resources such as unlicensed spectrum or an open deployment
of neutral small cells. In Section~\ref{sec:duopoly},
we describe a duopoly game involving two vertically differentiated
mmWave network service
providers, and compare their profits with and without resource
sharing, for simultaneous market entry and sequential market entry.
Finally, in Section~\ref{sec:conclusion}, we conclude with a discussion
of the implications of this work and areas of further research.

\section{Economic foundations}
\label{sec:economic}

We briefly summarize here the economic framework 
used in the rest of this paper. We define a 
network good and show how its demand 
is fundamentally different from demand for non-network goods, 
give equilbria and conditions for reaching critical 
mass in a market for a network good, 
explain the concept of compatibility, 
and describe a model of vertically differentiated network good.
For a more detailed overview of this area of economics, 
see~\cite{economides1996economics}.

\subsection{Network goods}
\label{subsec:networkgood}

In economics, a network good or service~\cite{economides1996economics}
is a product for which
the utility that a consumer gains from the product
varies with the number of other consumers of the product
(the \emph{size of the network}).
This effect on utility - which is called the
\emph{network externality} or the \emph{network effect} - 
may be direct or indirect.
The classic example of a direct network effect is the
telephone network, which is more valuable when the service
has more subscribers.
The classic example of an indirect effect
is the hardware-software model,
e.g. a consumer who purchases an Android smartphone
will benefit if other consumers also purchase
Android smartphones, because this will incentivize
the development of new and varied applications for
the Android platform. The network externality
may also be negative, for example, if an Internet
service provider becomes oversubscribed, its subscribers
will suffer from the congestion externality.

\begin{figure}[h]
\centering
\includegraphics[width=3in]{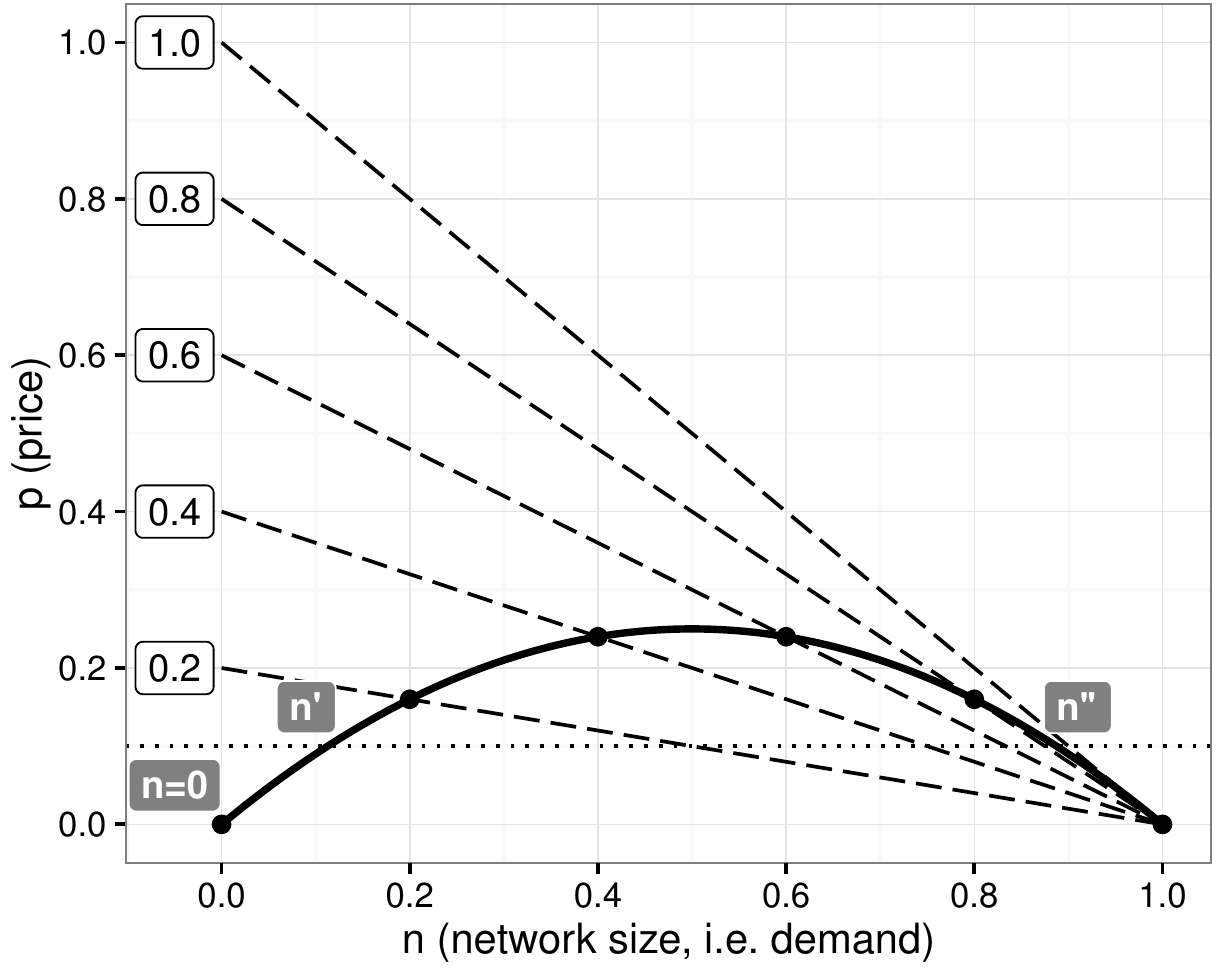}
\caption[]{An example of demand curves for a series of 
expected network sizes (dashed lines, each labeled with 
expected network size $n^e$), and 
the fulfilled expectations demand curve $p(n; n)$ (solid line), which 
is the collection of points where the demand curve for an expected
network of size $n^e$ intersects the vertical line at $n=n^e$. Three
equilibria for a perfectly competitive market with marginal cost $c=0.1$
are labeled in white text on a dark background, one at $n=0$ and two 
at the intersesctions of $p(n;n)$ and $p=c$ (dotted line).}
\label{fig:fulfilled-expect-example}
\end{figure}

A fundamental difference between a network good and one
with no network effects is the behavior of the demand 
curve, which describes the relationship
between two key quantities:
consumer demand for a good, $n$ (or equivalently, 
the number of units of the good that are sold i.e., network size), and
the price of the good, $p$.

Consider a set of consumers in a market for a non-network good. 
The total demand for the good, $n$, is normalized
so that $n=1$ when all consumers purchase
the good, and $n=0$ when no consumers purchase 
the good. A consumers' willingness to 
pay for the good is $\omega$, with 
$\omega$ varying
among the set of consumers up to $\hat{\omega}$ 
(different consumers
are willing to pay different prices
for an identical good).
A consumer of type $\omega$
is indifferent between purchasing the good 
or not when price $p = \omega$.
For $p > \hat{\omega}$, none of the consumers will purchase the 
good ($n=0$), because it is too expensive even for consumers
with the highest value of $\omega$. At $p=0$, 
all of the consumers will purchase the good ($n=1$). 
The demand curve, which indicates what portion 
of the consumers will purchase the good at a given 
price, has a negative slope for most kinds of non-network goods, 
because the quantity demanded $n$ (typically shown on the horizontal axis) 
increases as the price of the good $p$ (typically shown on the vertical axis)
decreases. 
Conversely, to increase demand for a 
typical non-network good, a producer of the good must reduce its price.
The dashed line labeled ``1.0'' 
in Fig.~\ref{fig:fulfilled-expect-example} shows a sample 
demand curve when $\omega$ is distributed uniformly
on the interval $[0,1]$.

Now let us consider a network good. We still assume 
heterogenous consumers, with type $\omega$ up to $\hat{\omega}$, 
but a consumer of type $\omega$ 
has willingness to pay $\omega h(n)$, 
where $n$ is the network size, 
and $h(n)$ is a network externalities 
function indicating how consumer utility
scales with $n$. For a positive network externality, 
$h(n)$ increases with $n$, and for a negative network 
externality $h(n)$ decreases with $n$.
A consumer purchasing the good at price $p$
gains utility $u(\omega, n, p) = \omega h(n) -p$.

Under these circumstances, a consumers' decision to 
purchase the good or not depends on how many units 
of the good they expect will be sold, i.e., the 
expected network size $n^e$. A consumer of type $\omega$ is indifferent
between purchasing the good or not when $p=\omega h(n^e)$.
We can draw a demand curve for any expected network size $n^e$.
For $p > \hat{\omega} h(n^e)$, none of the consumers will purchase the 
good ($n=0$). For $p=0$, 
all will purchase the good ($n=1$). 
At the point where the demand curve intersects the vertical 
line at $n=n^e$, i.e., when the demand for the good 
at a given price equals the expected network 
size, we say that consumers' expectations are fulfilled.

We can construct a series of such demand curves $p(n; n^e)$  
for different values of $n^e$.
Each curve gives the willingness to pay of the $n$th consumer
when the expected size of the network is $n^e$.
The collection of points at $n=n^e$, where the actual size of the
network and consumers' expectations regarding the size of the
network are the same,
then make up the fulfilled expectations demand curve, 
$p(n;n)$. Fig.~\ref{fig:fulfilled-expect-example} shows 
the fulfilled expectations demand curve $p(n;n)$  and
demand curves for selected values of $n^e$ 
when $h(n) = n$ and $\omega$ is distributed uniformly
on the interval $[0,1]$.

This fulfilled expectations demand curve
gives the size of the network that could be supported
at equilibrium for a given price, in the same way that the demand curve for 
a typical non-network good defines the demand that
can be supported at a given price.
We notice two key differences between the behavior of the 
demand curve for a network good and a non-network good.

First, the demand for a network good depends on the consumers'
expected utility, which in turn depends on the
expected network size $n^e$. However, the expected network
size depends on consumer demand,
creating a self-fulfilling expectation.
When many consumers expect the good to be unpopular,
they will not purchase the good, and the network size will be small. 

Second, we note that although a traditional demand
curve always slopes down, the fulfilled expectations demand
curve first increases with $n$, then decreases. 
That is, for goods with a positive network externality, 
$p(n;n)$ first increases
with $n$ due to the network externality, but eventually begins to
slope downward, as it becomes increasingly difficult to find
customers who have not yet purchased the good, but
have a high enough willingness to pay for it.
For a traditional good, to gain market share a producer has to \emph{reduce}
the price. For a network good, a producer can sometimes demand a higher 
price as more units of the good are sold, because
the consumers' utility increases with the number 
of units sold.

\subsection{Equilbria and critical mass}
\label{subsec:equilibria}

An important feature of a fulfilled expectations
demand curve is how it relates to the size of a network 
at equilibrium, and particularly, how it relates
to the \emph{critical mass} of the network: the smallest
network size that can be sustained in equilibrium~\cite{economides1995critical}.

In a monopoly, the producer of the good has no competition 
to drive down the price. The fulfilled expectations demand 
curve $p(n;n)$ defines the price that the producer can charge 
to sustain demand of size $n$, earning total revenue of $n p(n;n)$.
The producer will choose the
network size that maximizes its profits
$\pi(n, p, c)= n(p(n;n)-c)$, where $c$ is the marginal cost, i.e., the
cost to the producer of providing one unit of the 
good~\cite{economides1995critical}.

In a perfectly competitive market, with many producers
offering goods that are perfect substitutes, 
competing producers will drive down the price
until it is equal to marginal cost, i.e.
$p(n; n) = c$,
yielding three possible equilibria when $c$ is less
than the maximum of $p(n;n)$:
\begin{enumerate}
\item one stable equilibrium at $n = 0$ (representing a zero size network),
\item an unstable equilibrium for a network of size $n'$ at the first
(smaller $n$) intersection of the fulfilled expectations
demand curve $p(n; n)$ with the horizontal line
$p=c$, and
\item a stable equilibrium for a network of size $n''$ at
the second (larger $n$) intersection of $p(n;n)$ and the horizontal $p=c$.
\end{enumerate}
These are illustrated in Fig.~\ref{fig:fulfilled-expect-example}. 
When the line $p=c$ intersects $p(n;n)$ once, at
its maximum value, there is a stable equilibrium at that point
and one at $n=0$. When $c$ is greater than the maximum value
of $p(n;n)$, the producers would only be able to 
sell the good at a loss, so the only equilibrium will
be at $n=0$, and no producer will offer the good~\cite{economides1995critical}.

It is shown in~\cite{economides1995critical} 
that under perfect competition,
the critical mass is equal to the network size $n^0$ at which $p(n;n)$ is
maximized. At this point,
competing pressures on demand are perfectly balanced.
For network sizes between zero and $n'$,
there is ``downward pressure''
toward the first equilibrium at $n=0$, since
there are not enough consumers willing to
pay for the good at the lowest price
at which the producer is prepared to offer it.
When $n' < n < n''$,
there are more consumers willing
to pay price $c$, and
the service increases in value as more units are sold,
exerting ``upward pressure'' on the demand toward
the equilibrium at $n''$.
For $n > n''$,
there is again ``downward pressure'' on the demand
toward $n''$
because producers are trying to sell the good to the part of the
population with a low willingness to pay.

Because of these pressures on demand,
$n''$ has a strong stability
property, and $n'$ is highly unstable.
A producer entering a new market is interested in selling enough units
at a small network size
for the network to grow to at least $n'$,
since beyond that ``tipping point'' the upward pressure on
demand helps the network reach its non-zero stable equilibrium. 
The slope of the fulfilled
expectations demand curve $p(n;n)$ for small network sizes 
is very important, since it describes how easy it is 
for the network size to reach critical mass. 
When this slope is large, then for a given value of $c$, 
$n'$ occurs at a smaller network size, making it 
easier to reach critical mass and from there, the stable equilibrium
at $n''$.

\subsection{Compatibility}
\label{subsec:compatibility}

In the previous section, we model a consumer's willingness to 
pay for good $i$ as $\omega h(n_i)$, where $h(n_i)$ is the network 
externalities function and $n_i$ is the number of consumers
who have purchased the good. In this model, $h(n_i)$ is not affected 
by the number of units sold of any other good.
Now we consider a market where 
producers may choose to make their goods 
\emph{compatible}~\cite{economides1998equilibrium,katz1985network}.
When two network goods are compatible, then the total network
effect for a consumer of either good is based on
the sum network size of both goods, so the network externalities 
function for good $i$ is evaluated using the total network size 
for all the goods: 
$h( \sum_{j \in I} n_{j})$, where
$I$ is a set of firms producing compatible goods and $i\in I$.

A firm producing
a network good has conflicting incentives for and
against compatibility:
\begin{itemize}
\item Positive network externalities: A firm that makes its
product \emph{compatible} increases its value to consumers, 
since the argument to $h(\cdot)$ is greater.
\item Market power: A firm that chooses to make its product
\emph{incompatible} reduces the value of its competitors' product, 
so it avoids losing market share, and can charge higher prices.
\end{itemize}
Compatibility is often used to model a firm's
choice to use a proprietary technical standard or a common industry standard.
For example, the developers of a word processing application 
might choose to use a proprietary file format so that all consumers
who need to open these files must purchase their software, 
or they might choose to use an open standard so that their
users can share the files produced with their software 
with users of other word processors. 

\subsection{Vertical differentiation}
\label{subsec:differentiation}

In our previous model, when two network goods are compatible, consumers
prefer them equally, since the value of the network 
externalities function is the same for both.
This can shift demand from one producer to another. 
A producer may try to disinguish itself
from competitors by improving 
the value of its good in other ways
(not by increasing $n$). For example, consumers of 
Android-based smartphones benefit from the network effects
due to consumers of all Android-compatible smartphones.
However, 
a firm that produces Android phones can distinguish 
itself in the market by selling handsets with better 
hardware specifications than its competitors'.

In Section~\ref{sec:duopoly} of this paper, where we consider
service providers that are already well established 
in the market, we use a model 
of consumer utility described in~\cite{baake2001vertical} which 
includes vertical differentiation. In this model, a consumer's willingness to pay
for good $i$ is $\omega q_i + q_i h( \sum_{j \in I} n_{j})$, where
$I$ is a set of firms producing compatible goods, $i\in I$, 
and $q_i$ is a scaling factor that represents aspects of the good's
quality that are \emph{not} a function 
of the network size. Firms that produce compatible goods 
can distinguish themselves from one another
by choosing different quality levels. However, a firm that 
chooses to produce a higher-quality good also has higher
marginal costs. Where $c$ was a constant marginal 
cost in previous sections, now we scale marginal cost
according to the quality level, so that the cost to the producer of 
producing one unit of good $i$ at quality level $q_i$ 
is $q_i$, and its profits are $\pi_i(q_i, n_i, p_i) = n_i p_i - q_i n_i$.

\subsection{Application to mmWave network service}

Given this economic framework, we are interested in modeling 
mmWave network service as a network good, to better understand:
\begin{itemize}
\item What kind of network effects apply to mmWave network service?
What is the relative benefit to a consumer of subscribing
to a large service provider?
(Section~\ref{sec:network})
\item What is the behavior of the fulfilled expectations demand
curve for mmWave network service? How difficult is it for a service
provider just entering the market to reach critical mass? 
Do open resources help a service provider reach critical mass?
(Section~\ref{sec:demand})
\item Under what conditions will mmWave network service providers want 
to share resources? Is it desirable for a regulator to 
enforce resource sharing?
(Section~\ref{sec:duopoly})
\end{itemize}

To answer these questions, we must connect the economic models
described in this section to an accurate technical model of 
mmWave cellular systems, described in Section~\ref{sec:technical}.

\section{Technical benefits of resource sharing}
\label{sec:technical}

In this section, we describe the system model of the mmWave
network used in the rest of the paper. We outline 
the technical benefits of resource sharing between 
identical service providers, confirming some of the results 
of~\cite{andrews-sharing,european-sharing,matiamag,matia}.
We also describe the sharing gains achieved by service providers
that are asymmetric with respect to number of subscribers,
spectrum holdings, and base station deployments.
The simulations of this section will be used in Section~\ref{sec:network}
and Section~\ref{sec:demand}
to devise an economic model for mmWave resource sharing.

\subsection{mmWave System Model}

We consider a system with multiple mmWave network service
providers (NSPs) operating in the 73 GHz band.
A service provider $i \in\{1,\dots,I\}$
has bandwidth $W_i$, a set of base stations (BSs)
distributed in the network area using a
homogeneous Poisson Point Process (hPPP) with intensity $\lambda^B_i$,
and a set of user equipment (UEs) whose locations are modeled by an
independent hPPP with intensity $\lambda^U_i$.

Both BSs and UEs use antenna arrays for directional beamforming.
For the sake of tractability, we approximate the actual array patterns
using a simplified pattern as in \cite{heath-mm,andrews-antenna}.
Let $G(\phi)$ denote the simplified
antenna directivity pattern depicted in Fig. \ref{pattern},
where $M$ is the main lobe power gain, $m$ is the back lobe
gain and $\theta$ is the beamwidth of the main lobe.
In general, $m$ and $M$ are proportional
to the number of antennas in the array and $M/m$ depends
on the type of the array. Furthermore,
$\theta$ is inversely proportional to the number of
antennas, i.e., the greater the number of antennas,
the more beam directionality. We let $G^B(\phi)$
(which is parameterized by $M^B$, $m^B$, and $\theta^B$) be
the antenna pattern of the BS, and $G^U(\phi)$
(which is parameterized by $M^U$, $m^U$, and $\theta^U$) be the antenna pattern
of the UE.

\makeatletter
\if@twocolumn%
    \begin{figure}[!htb]
    \centering
    \includegraphics[width=2.8in]{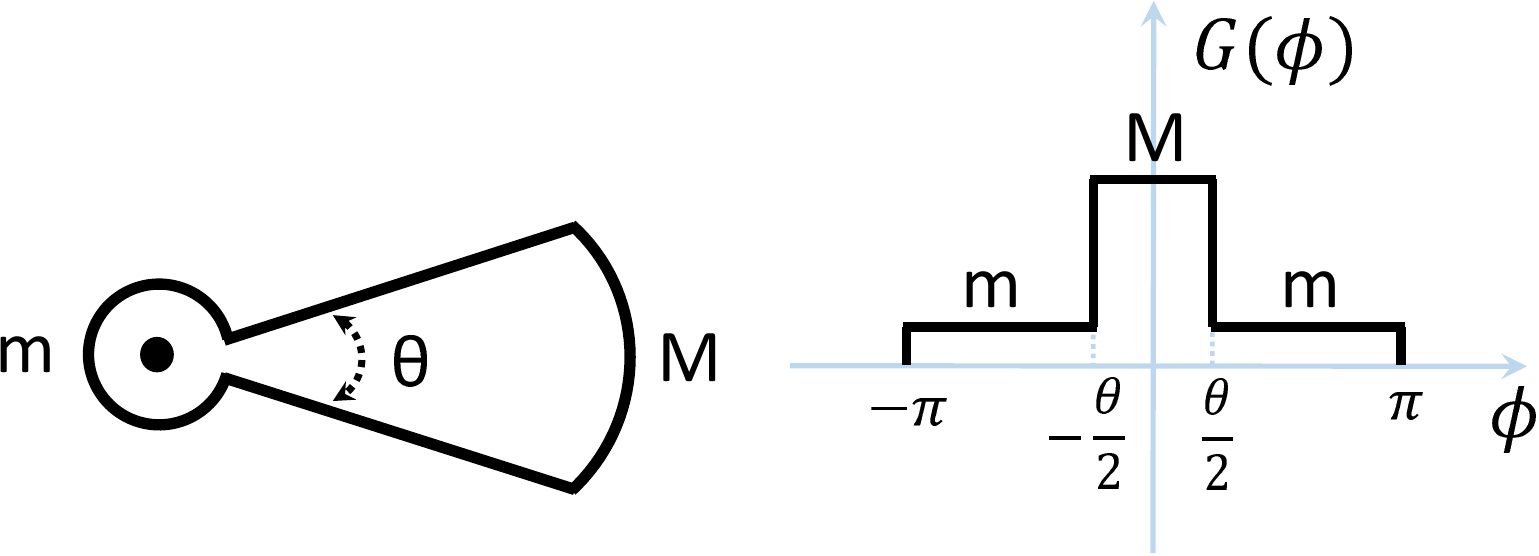}
    \caption[]{Simplified antenna pattern with main lobe $M$, back lobe $m$ and beamwidth $\theta$.}
     \label{pattern}
    \end{figure}%

    \begin{figure}[!htb]
    \centering
    \includegraphics[width=2.3in]{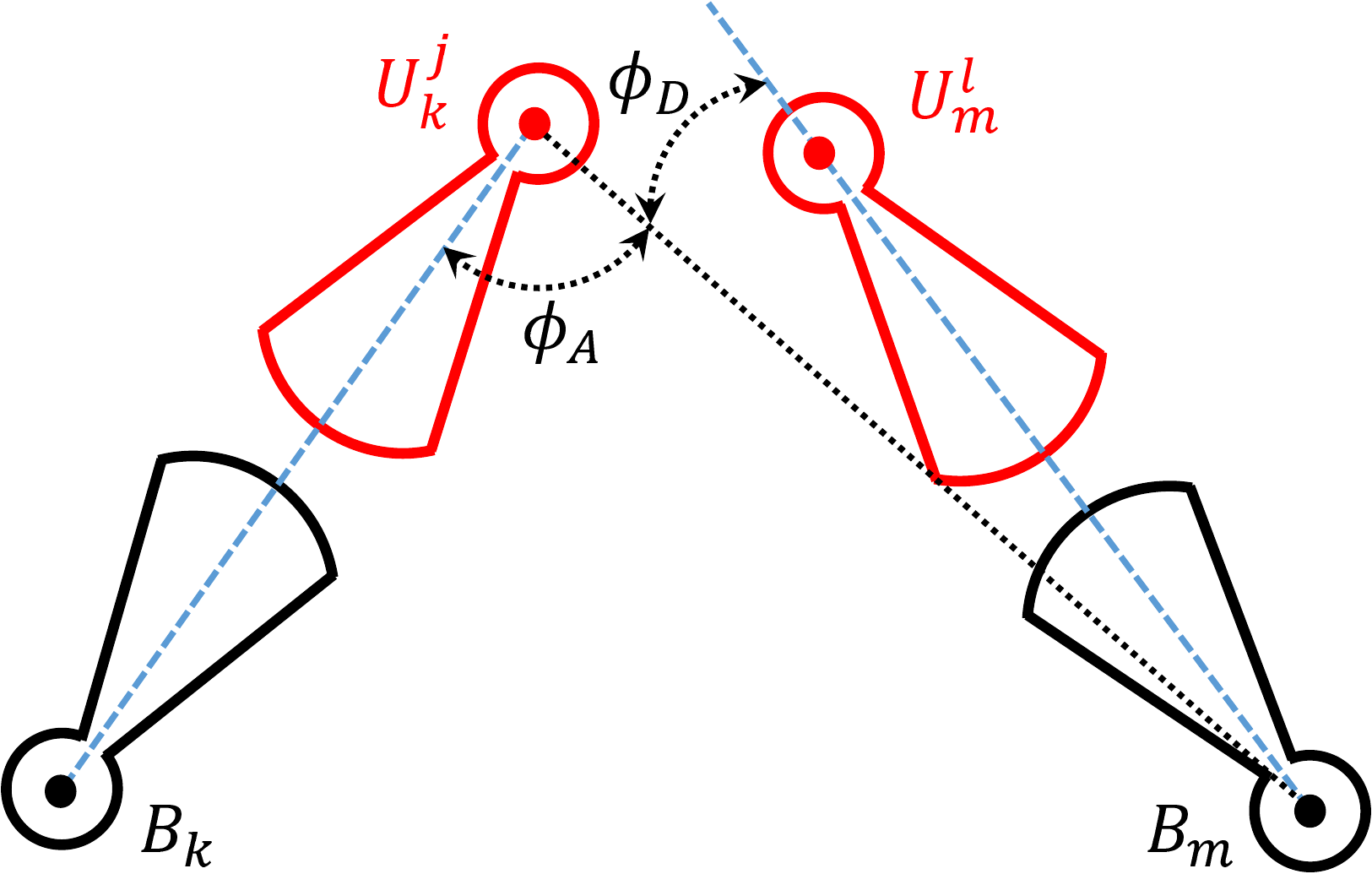}
    \caption[]{Intercell interference from base station $B_m$ received at the
    UE $U^j_k$, which is associated with neighboring base station $B_k$.
    $\phi_D$ is the departure angle from $B_m$ and $\phi_A$ is the arrival angle at $U^j_k$.}
    \label{interference}
    \end{figure}%

\else
    \begin{figure}[!htb]
        \centering
        \begin{minipage}{.46\textwidth}
    \includegraphics[scale=0.45]{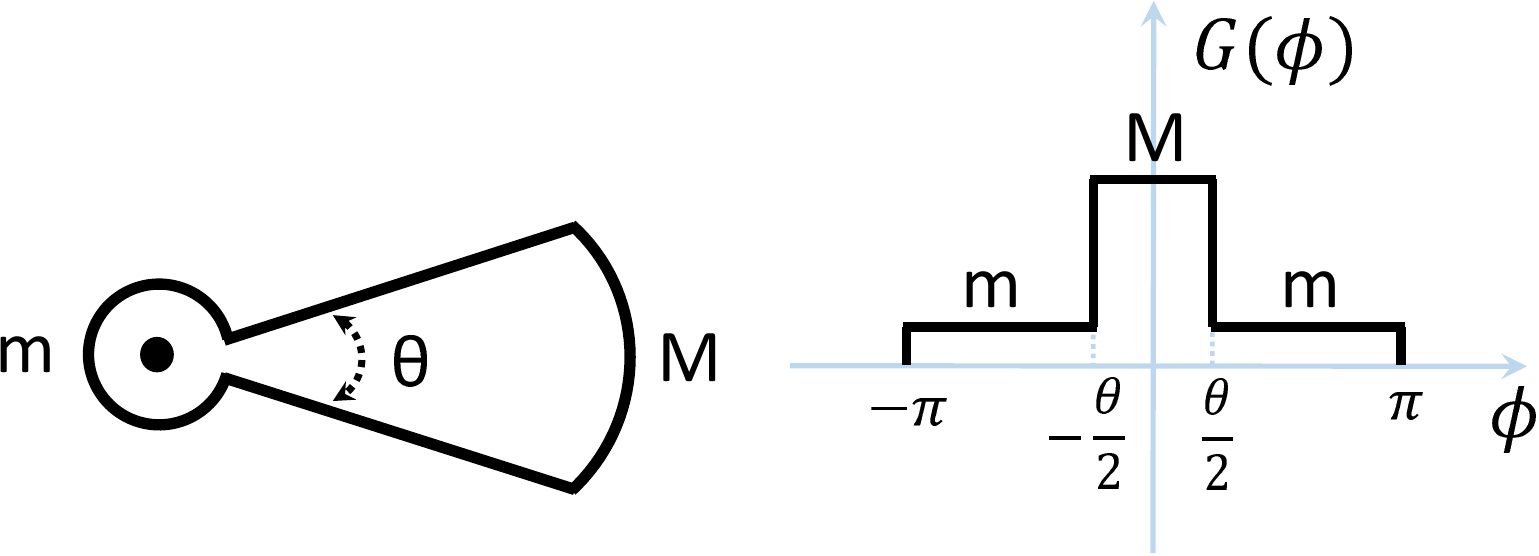}
    \caption[]{Simplified antenna pattern with main lobe $M$, back lobe $m$ and beamwidth $\theta$.}
     \label{pattern}
        \end{minipage}%
        \hfill
        \begin{minipage}{0.52\textwidth}
        \centering
    \includegraphics[scale=0.32]{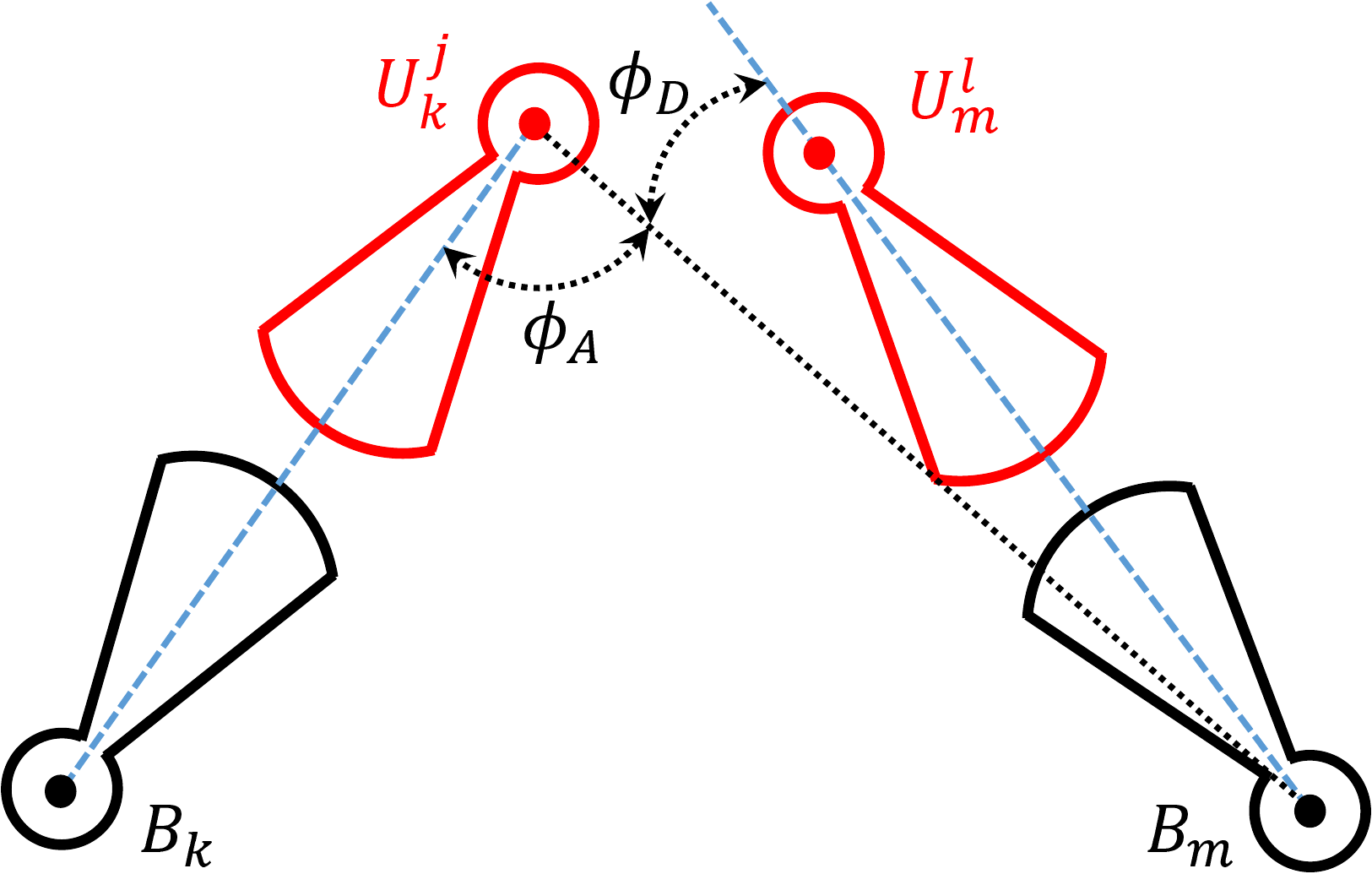}
    \caption[]{Intercell interference from base station $B_m$ received at the
    UE $U^j_k$, which is associated with neighboring base station $B_k$.
    $\phi_D$ is the departure angle from $B_m$ and $\phi_A$ is the arrival angle at $U^j_k$.}
    \label{interference}
        \end{minipage}
    \end{figure}
\fi
\makeatother

We consider a time-slotted downlink of a mmWave cellular system.
The channel model we use includes path loss, shadowing, outage, 
and small scale fading.
For path loss, shadowing,
and outage, line of sight (LOS), and NLOS probability distributions,
we use models adopted from \cite{mustafa-channel}.
We assume Rayleigh block fading.
Finally, the data rate is modeled as
\begin{equation}
R=(1-\alpha)W\log_2 \Bigg(1+\beta \frac{PG^U(0)G^B(0)H}{N_fN_0W+I}\Bigg),
\label{rate-model}
\end{equation}
where $\alpha$ and $\beta$ (which are specified in Section~\ref{subsec:performance}) 
are overhead and loss factors, respectively,
and are introduced to fit a specific physical layer to the
Shannon capacity curve.
Furthermore, $P$ is the BS transmit power,
$H$ is the channel power gain derived from the model
discussed above, and $N_f$, $N_0$, $W$ and $I$ are UE noise figure,
noise power spectral density, bandwidth, and interference power, respectively.
We assume perfect beam alignment between BS and UE within a cell,
therefore the antenna power gain (link directionality) is
$G^U(0)G^B(0)=M^UM^B$.
The SINR of a UE is defined as $\frac{PG^U(0)G^B(0)H}{N_fN_0W+I}$.

Let $B_k$ denote the BS of cell $k\in\{1,\dots,K\}$, where $K$ is the
total number of cells. Also, let $U_k^j$ denote the UE $j\in\{1,\dots,N_k\}$
of cell $k$, where $N_k$ is the total number of the UEs in cell $k$. We do not
introduce any intercell coordination to manage interference. The power of intercell interference
from base station $B_m$ received at UE $U_k^j$
depends on the beamwidth $\theta$, the
arrival angle $\phi_A$ at $U_m^j$, and the departure angle
$\phi_D$ from $B_k$, as shown in Fig.~\ref{interference}.
In the scenario illustrated we have $I=PG^U(\phi_A)G^B(\phi_D)H_m^{k,j}$, 
where $H_m^{k,j}$ here
is the channel gain between $B_m$ and $U_k^j$.
While the probability of strong
intercell interference is low due to the beam directionality,
interference still exists.
There are four main factors affecting intercell interference:
\begin{itemize}
 \item{\textbf{Frequency:}}
    The interference signal is weaker at high frequencies due to path loss \cite{andrews-sharing}.
 \item{\textbf{Antenna Pattern:}}
    Antenna pattern parameters, especially the beamwidth ($\theta$ in Fig.~\ref{pattern}), directly affect the intercell interference. A larger
    beamwidth results in stronger intercell interference \cite{heath-mm,andrews-sharing}.
 \item{\textbf{UE and BS density:}}
    In mmWave networks, strong intercell interference 
    occurs infrequently in sparse networks~\cite{heath-mm}.
    This is also true for conventional cellular networks (like LTE), but with the greater beam directionality used in mmWave networks, the network becomes interference limited for much higher UE and BS densities than in conventional networks. However, for ultra-dense deployments, intercell interference can be dominant even with high beam directionalities if there is no coordination~\cite{heath-mm,european-sharing}.
 \item{\textbf{Bandwidth:}}
    When the network bandwidth is large, noise power becomes the dominant factor
    and interference is less important~\cite{mustafa-channel}.
\end{itemize}

We consider two kinds of intercell interference.
Intra-NSP interference comes from transmissions of neighboring
BSs to UEs of the same NSP.
Inter-NSP interference comes from transmissions of neighboring
BSs to UEs of a different NSP
on the same frequency, and occurs only
when the service providers use spectrum on a non-exclusive basis.

To model interference, we consider four kinds of resource use:
\begin{enumerate}
\item \textbf{Exclusive Spectrum, Exclusive BSs (No Sharing)}:
Each NSP works
independently on its own part of the spectrum, with its own BSs.
Obviously, there is no inter-NSP interference.
Each UE associates with the closest BS
belonging to its own NSP.
\item \textbf{Exclusive Spectrum, Non-Exclusive BSs (BS Sharing Only)}: 
A UE associates with the closest BS,
regardless of which NSP it belongs to.
As a result, the average distance between a UE and its serving BS
is reduced, which leads to a higher received
signal power. However, this also increases the
intra-NSP interference power relative to the no sharing case,
because the distance
between the interferer BSs and a UE shrinks along with cell radius.
There is no inter-NSP interference.
\item \textbf{Non-Exclusive Spectrum, Exclusive BSs (Spectrum Sharing Only)}:
NSPs use spectrum on a non-exclusive basis.
Besides for the increase in
noise power due to larger bandwidth, inter-NSP interference also occurs in
addition to intra-NSP interference,
and these together decrease the SINR of UEs
compared to the previous cases, but may still increase the rate.
\item \textbf{Non-Exclusive Spectrum, Non-Exclusive BSs (Full Sharing)}:
NSPs use both spectrum and BSs on a non-exclusive basis. This case is a
combination of Case 2 and Case 3.
\end{enumerate}

\subsection{Scheduling}

Early work on resource sharing in mmWave
networks~\cite{andrews-sharing,european-sharing,matiamag,matia}
has focused on signal propagation
and interference effects in networks with shared resources.
To approximate data rate at a UE, these papers divide
the link capacity as determined by the UE's average  
SINR by the total number of UEs
in the cell. In a realistic network with opportunistic
scheduling, however, a UE may achieve a higher data rate
than its average SINR would suggest, because it is scheduled
with higher priority in time slots when its SINR is high.
This scheduling gain increases with the number of UEs.
For an economic analysis we need to accurately model
how consumers' utility scales with all aspects of
network size, including the number of subscribers,
so our model must include this scheduling gain.

We adopt a modified scheduler based on
the multicell temporal fair opportunistic scheduler proposed
in~\cite{shahram-scheduler}. 
We use only the first stage (UE nomination stage) of this scheduler, 
since there is no coordination among the BSs. We expect the difference 
from the two-stage scheduler with coordination to be neglible, since
intercell interference is limited by the directional
nature of the transmissions in mmWave networks~\cite{matiamag}.
Thus each BS runs the scheduler and selects a UE
independently, without considering intercell interference
and making scheduling decisions based only on the signal
to noise ratio of the UEs in each time slot.

\subsection{Performance Evaluation}
\label{subsec:performance}

In order to establish benefits of resouce sharing 
from a technical perspective, we present simulation results of a mmWave
network with two NSPs ($i\in\{1,2\}$) operating in the 73 GHz band.
We fix BS transmit power ($P$) and loss factor ($\beta$) as 30 dBm and 0.5,
respectively, as in \cite{mustafa-channel}. Table~\ref{sim-param} shows
the specific parameters we use in the simulations. We assume two
different cases: \emph{symmetric NSPs} and
\emph{asymmetric NSPs}. In the symmetric case,
NSPs are identical in terms of network resource
(BS density, bandwidth) and UE density. In the 
asymmetric case, one of the NSPs has more network
resources and UE density than the other. Parameters such as UE and
BS densities ($\lambda^U_i$, $\lambda^B_i$), and bandwidth ($W_i$)
will be specified for each NSP separately.

\begin{table}[ht]
\caption{Network parameters}
\centering
\begin{tabular}{llr}
\toprule \textbf{Parameter} & \textbf{Value} \\
\midrule
Frequency                                       &  73 GHz \\
Total bandwidth ($W_1+W_2$)                     &  1 GHz \\
Total BS density ($\lambda^B_1$+$\lambda^B_2$)  &  100 BSs/$\text{km}^2$ \\
Total BS density ($\lambda^U_1$+$\lambda^U_2$)  &  500 UEs/$\text{km}^2$ \\
BS transmit power $P$                           &  30 dBm \\
BS antenna model ($M^B$,$m^B$,$\theta^B$)       & (20 dB, -10 dB, 5\degree)\\
UE antenna model ($M^U$,$m^U$,$\theta^U$)       & (10 dB, -10 dB, 30\degree)\\
Rate model ($\alpha$, $\beta$)                  & (0.2, 0.5)\\
UE noise figure $N_f$                           & 7 dB\\
Noise PSD $N_0$                                 & -174 dBm/Hz\\
Simulation duration $T$                         & $10^5$ slots\\
\bottomrule
\end{tabular}
\label{sim-param}
\end{table}

\subsubsection{Symmetric NSPs}
\label{subsec:symmetric}

\begin{figure}[t]
\centering
\includegraphics[width=3.45in]{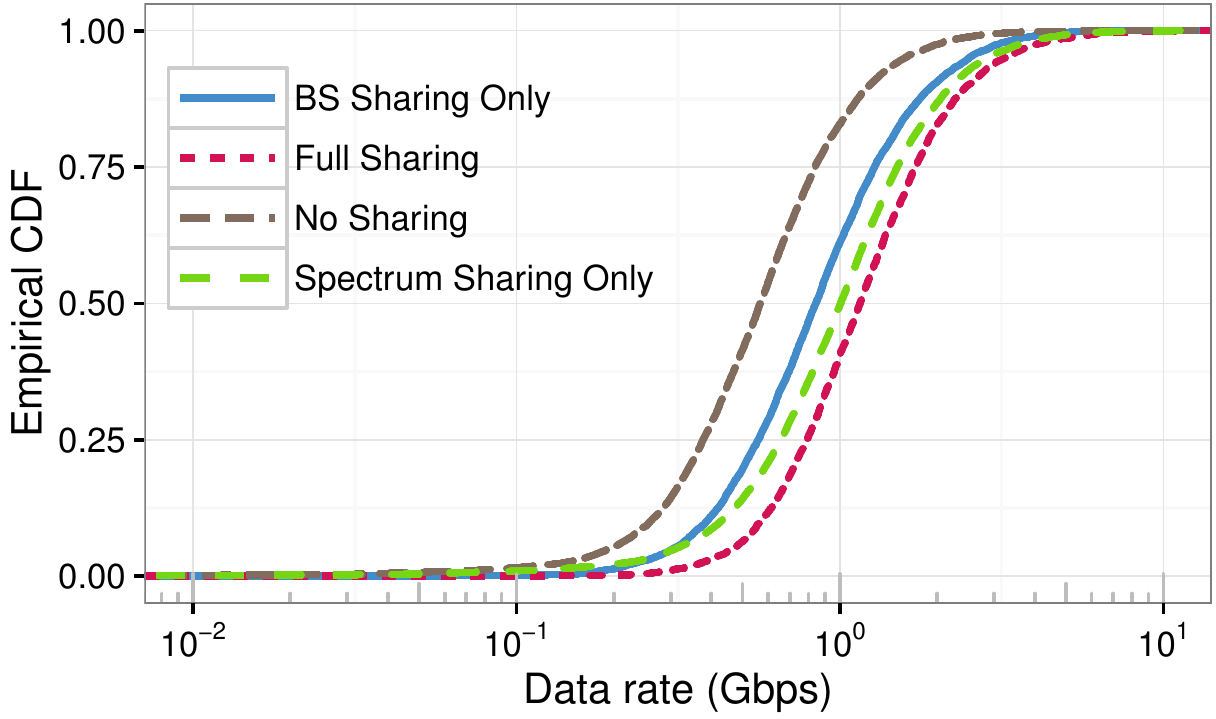}
\caption[]{There is a positive effect on
UE rate when symmetric NSPs pool their BS
and spectrum resources.}
\label{fig:symmetric}
\end{figure}

Fig.~\ref{fig:symmetric} shows the cumulative distribution function (CDF)
of the UE rate, for symmetric NSPs, 
each with 500 MHz of 73 GHz spectrum licensed for exclusive use, 50 BSs,
and 250 UEs in a single square kilometer. All UEs benefit the most
from when the NSPs pool both spectrum
and BSs. However, both spectrum sharing alone
and BS sharing alone improve UE rate relative to
the case where no resources are shared.
BS sharing has a greater effect on rate
than spectrum sharing for UEs outside the coverage range
or with a poor signal quality.
For UEs in outage, BS sharing improves
coverage probability. For UEs with a low SNR,
there is little benefit to adding bandwidth because
they are in a power-limited regime.
UEs with a good signal
quality (high SNR) are bandwidth-limited and benefit more
from spectrum sharing than from BS sharing.
At this BS density (100 BSs total per square
kilometer), the effects of interference are neglible
due to the directional nature of the transmissions,
so there is no negative effect due to spectrum
sharing without coordination.

When two NSPs pool their BS and
spectrum resources, they offer UEs a higher
data rate. This is consistent with the results described
in~\cite{andrews-sharing,european-sharing,matiamag,matia}.
However, the early work on mmWave resource sharing
in~\cite{andrews-sharing,european-sharing,matiamag,matia}
does not consider NSPs, which 
we address next.

\subsubsection{Asymmetric NSPs}
\begin{figure*}[ht]
\centering
\includegraphics[width=0.92\linewidth]{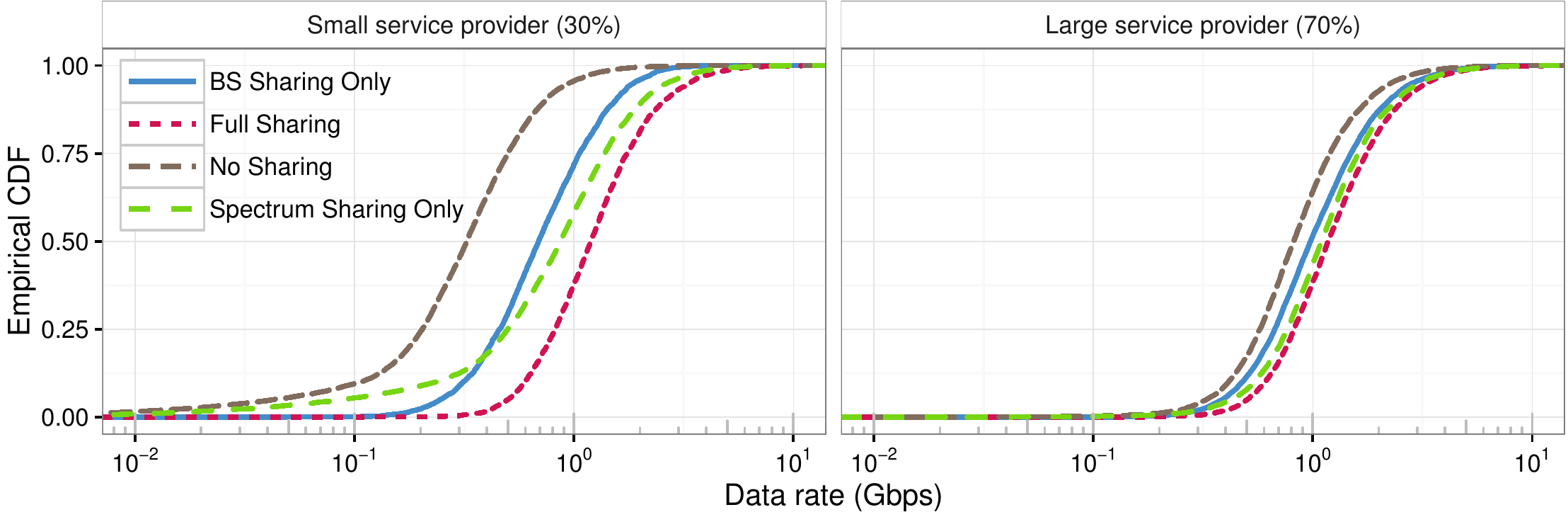}
\caption[]{When asymmetric NSPs share spectrum
and BS resources, they provide a higher
quality of service, but are no longer able to differentiate
themselves on the data rates they offer.}
\label{fig:asymmetric}
\end{figure*}

Fig.~\ref{fig:asymmetric} shows the CDF of UE rate for
consumers of two \emph{asymmetric} NSPs operating in the same
geographic area. The larger NSP has $70\%$ of
resources and subscribers: 700 MHz of 73 GHz spectrum licensed for exclusive use,
70 BSs, and 350 UEs in a single square kilometer.
The smaller NSP has $30\%$ of
resources and subscribers: 300 MHz of spectrum, 30 BSs,
and 150 UEs.

When there is no resource sharing, the larger NSP gains
market power by offering
its subscribers a higher rate than competing services.
With fully shared resources, however, their
subscribers' rate distributions are identical and the 
services are perfect substitutes.
Sharing increases the value
of the service that both NSPs offer,
but eliminates the ability of the larger NSP
to distinguish itself in the market
by offering higher rates.

With only spectrum sharing or only BS sharing,
the larger NSP retains some ability to differentiate itself
in the market by offering higher rates.
However, even under these circumstances,
the small NSP enjoys greater
relative gains than the large NSP 
from sharing of any kind (compared to no sharing),
while the large NSP contributes more resources.
The small NSP especially benefits from
full sharing or BS sharing by increasing the coverage
probability of UEs that were in outage. This gives it an extra
competitive edge in the market among consumers
who consider stability and coverage probability
as a primary factor in choosing their NSP.

\section{mmWave Service as a Network Good}
\label{sec:network}

In the simulations of Section~\ref{sec:technical}, we 
assumed a fixed number of UEs and resources. 
Now we model mmWave network service
as a \emph{network good} (as explained in Section~\ref{sec:economic}), 
with varying demand and resources.
Subscribers benefit from an
indirect positive network externality: a
large wireless service provider with more
subscribers will build a denser
deployment of BSs, and purchase more spectrum.
(Given large available bandwidth at mmWave frequencies,
we expect it
will be feasible for NSPs to acquire more spectrum at will.)
We describe the impact of increasing network size 
on consumers' data rates, using the simulation
described in Section~\ref{sec:technical}.

The network size $n$ of a mmWave network is defined differently depending 
on the scenario:
\begin{itemize}
\item \textbf{No open resources}: In this scenario, an
NSP scales its spectrum licenses 
and BSs according to the number of subscribers 
it has. Network size, $n$, is the normalized demand for the service, 
but is also a scaling factor on the BS 
density ($ \lambda^{B}$) and bandwidth ($W$) of the NSP.
\item \textbf{Open BS deployment}: In this scenario, 
there is a preexisting deployment of neutral
small cells, operated by a coalition of service providers
or by a third party (as suggested in~\cite{industry-5g}).
These cells have an open association policy, and will serve
UEs of any NSP. Network size, $n$, refers to demand for the service and 
is also a scaling factor on the bandwidth of the NSP, $W$, but 
the BS density of the NSP is constant and equal 
to the size of the ``open''
deployment ($\lambda_\text{max}^{B}$) for all values of $n$.
\item \textbf{Open spectrum}: In this scenario, 
spectrum is unlicensed and may be used by any NSP. 
Here, $n$ refers to demand for the service and 
is also a scaling factor on the BS density of the NSP, $ \lambda^{B}$. 
However, the bandwidth of the NSP 
is constant and equal to the full unlicensed bandwidth
$W_{\text{max}}$ for all values of $n$.
\end{itemize}
Note that the use of ``open'' resources is not the 
same as sharing resources acquired by individual NSPs. Open resources
are fixed in size and available to all NSPs. 
Shared resources are also available to participating NSPs, 
but their size varies according to the network size of the NSPs.
Table~\ref{tab:networksize} enumerates the BS density and bandwidth 
used by NSP $i$ in all three scenarios, when there is no sharing between 
individual NSPs and when resources are shared among NSPs in $I$, 
with $i \in I$. Resources that are used by NSP $i$ on a non-exclusive basis
are in bold font.

\makeatletter
\if@twocolumn%

    \begin{table}[h]
    \caption{Scaling resources with network size}
    \centering
    \begin{tabular}{p{1.4cm}lp{1.6cm}p{1.6cm}p{1.6cm}}
    \toprule & & \textbf{No open resources} & \textbf{Open BS deployment} & \textbf{Open spectrum} \\
    \midrule
    No sharing     &    $\lambda_i^{B}$   & $ n_i \lambda_{\text{max}}^{B}$  & $ \bm{\lambda_{\text{max}}^{B}}$  &  $ n_i \lambda_{\text{max}}^{B}$ \\
           &   $W_i$  &  $n_i W_{\text{max}}$    & $ n_i W_{\text{max}}$  & $ \bm{W_{\text{max}}}$  \\ \midrule
    Sharing  &  $\lambda_i^{B}$ & $ \bm{\sum_{j \in I} n_{j} \lambda_{\text{max}}^{B} }$  & $ \bm{\lambda_{\text{max}}^{B}}$  &  $ \bm{\sum_{j \in I} n_{j} \lambda_{\text{max}}^{B}}$ \\
            & $W_i$   & $\bm{\sum_{j \in I} n_{j} W_{\text{max}} }$    & $\bm{\sum_{j \in I} n_{j} W_{\text{max}}}$  & $\bm{W_{\text{max}}}$  \\
    \bottomrule
    \end{tabular}
    \label{tab:networksize}
    \end{table}
\else
    \begin{table*}[ht]
    \caption{Scaling resources with network size}
    \centering
    \begin{tabular}{llll}
    \toprule & \textbf{No open resources} & \textbf{Open BS deployment} & \textbf{Open spectrum} \\
    \midrule
    No sharing            & $ \lambda_i^{B} = n_i \lambda_{\text{max}}^{B}$  & $ \lambda_i^{B} = \bm{\lambda_{\text{max}}^{B}}$  &  $ \lambda_i^{B} = n_i \lambda_{\text{max}}^{B}$ \\
                & $W_i = n_i W_{\text{max}}$    & $W_i = n_i W_{\text{max}}$  & $W_i =\bm{W_{\text{max}}}$  \\
    Sharing  & $ \lambda_i^{B} = \bm{\sum_{j \in I} n_{j} \lambda_{\text{max}}^{B} }$  & $ \lambda_i^{B} = \bm{\lambda_{\text{max}}^{B}}$  &  $ \lambda_i^{B} = \bm{\sum_{j \in I} n_{j} \lambda_{\text{max}}^{B}}$ \\
                & $W_i =  \bm{\sum_{j \in I} n_{j} W_{\text{max}} }$    & $W_i = \bm{\sum_{j \in I} n_{j} W_{\text{max}}}$  & $W_i = \bm{W_{\text{max}}}$  \\
    \bottomrule
    \end{tabular}
    \label{tab:networksize}
    \end{table*}
\fi
\makeatother

Depending on the scenario, the total network externality $h(n)$ is the sum
of the network effects associated with up to three aspects of 
$n$. We separately quantify the contribution
to $h(n)$ of each:
\begin{enumerate}
\item \textbf{BS density}: We simulate a network
of increasing BS density, with bandwidth constant
at 1~GHz and ratio of UEs to BSs constant at 5 UEs
per BS. To separately quantify the effect of interference,
we also compute the UE rate based on SNR (neglecting the effect
of interference) and compare this to the actual UE rate. These
results are shown in Fig.~\ref{fig:ratebs}.
\item \textbf{Bandwidth}: We simulate a network
of increasing bandwidth, while keeping BS density constant
at a moderate value of 100 BSs per square kilometer,
and UE density constant at 500 UEs per square kilometer. These 
results are shown in Fig.~\ref{fig:ratebw}.
\item \textbf{UE density}: We simulate a network of increasing
UE density, while keeping bandwidth constant at 1~GHz and
BS density constant at 100 BSs per square kilometer.
To quantify the scheduling gain, we also simulate this network
with a round robin scheduler, and compare this to UE rate
with a partially opportunistic scheduler. These
results are shown in Fig.~\ref{fig:rateuser}.
\end{enumerate}

We assume that consumers decide whether or not to subscribe
to a mmWave network based on its fifth percentile rates.
This is supported by research on human behavior,
which suggests that service reliability is rated more highly
than overall connection quality in perceived quality of mobile
value-added services~\cite{Kuo2009887}. We take fifth percentile
rate as a proxy for service reliability.

\makeatletter
\if@twocolumn%
    \begin{figure}[!t]
        \centering
    \includegraphics[width=2.6in]{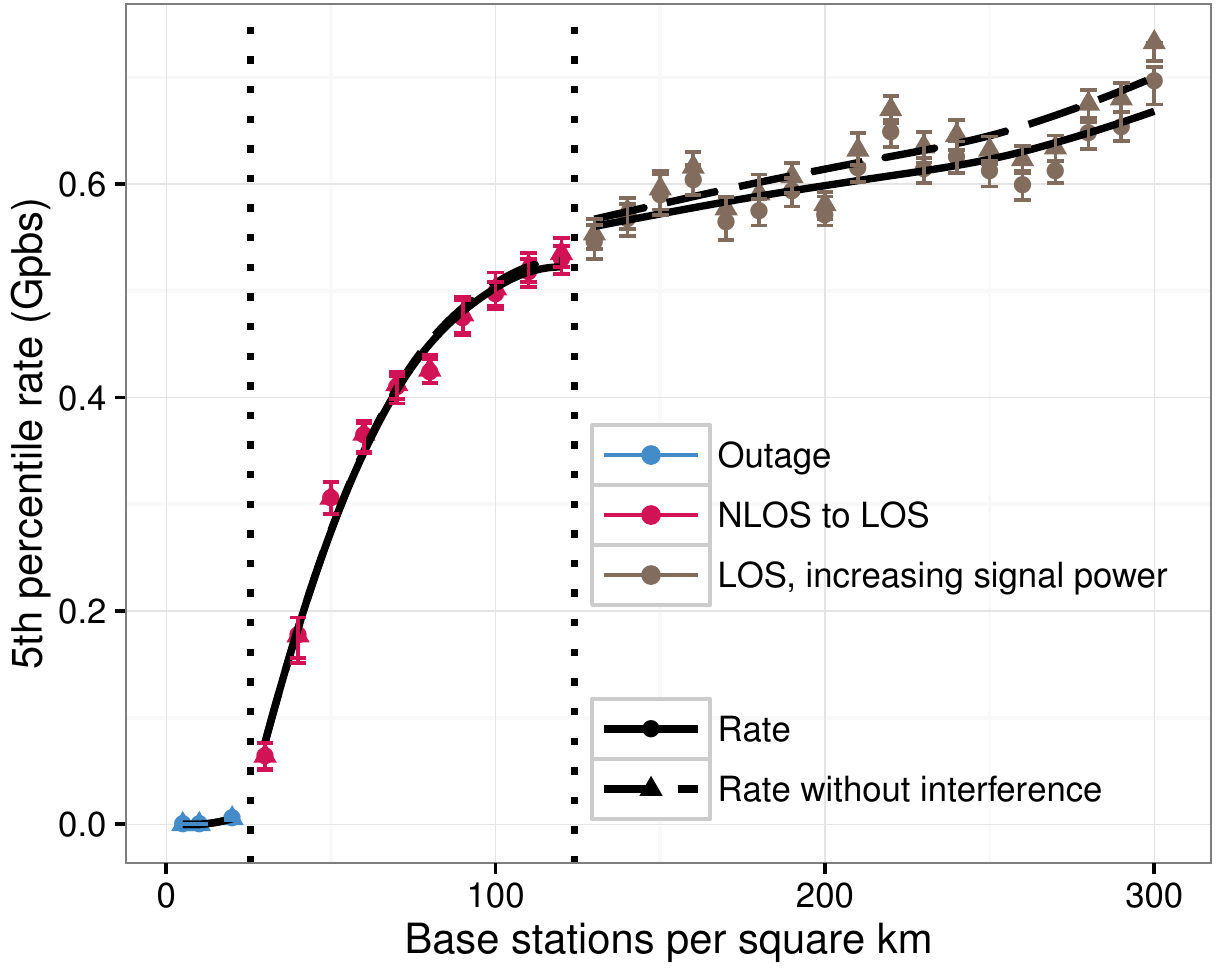}
    \caption[]{Effect of increasing BS density on fifth
    percentile rate as bandwidth is held constant at 1 GHz and UE
    density per BS is held constant at 5 UEs/BS.
    Error bars show 95\% bootstrap confidence intervals.}
    \label{fig:ratebs}
    \end{figure}

    \begin{figure}[!t]
        \centering
    \includegraphics[width=2.6in]{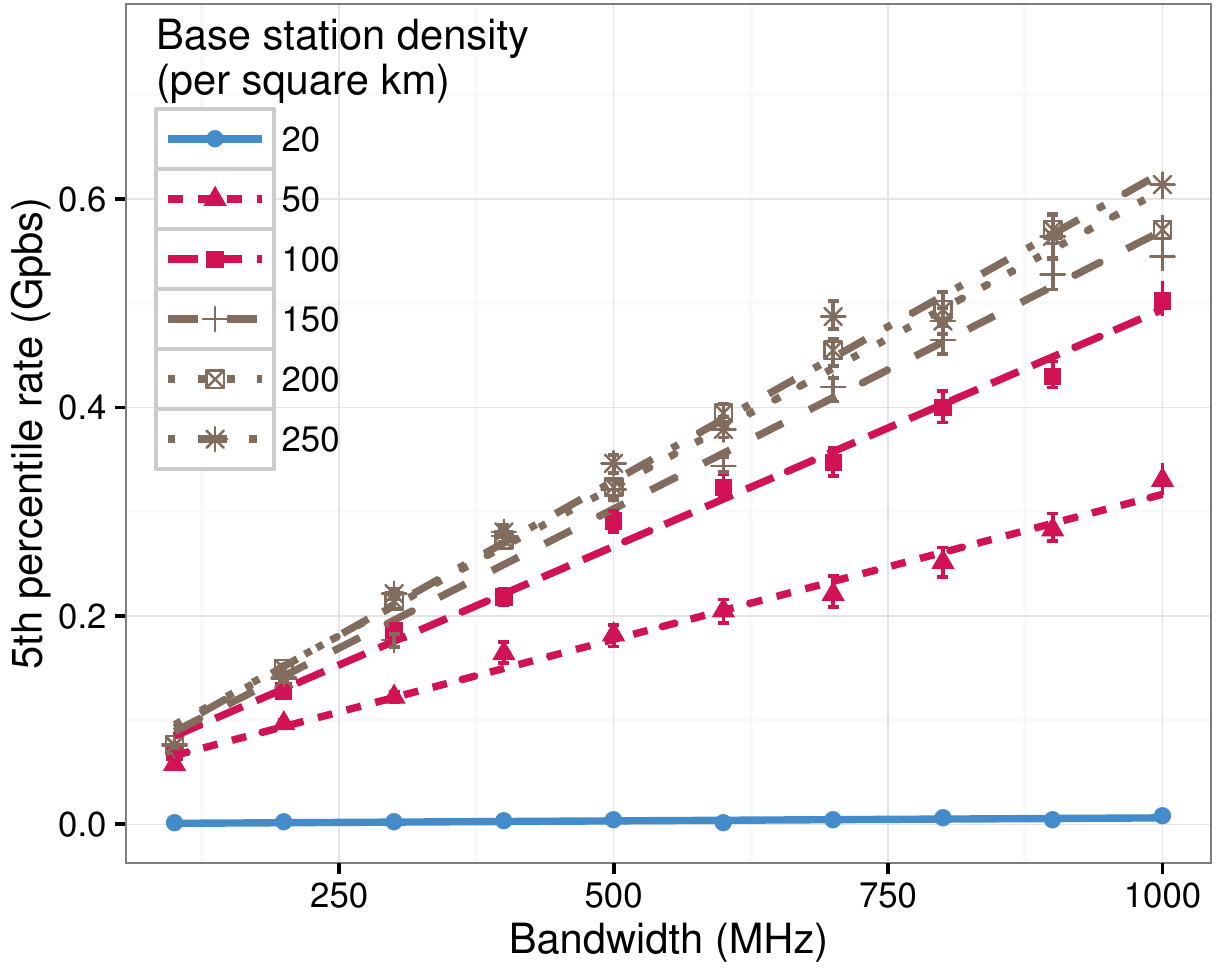}
    \caption[]{Effect of increasing bandwidth on fifth percentile
    rate for different BS densities, with UE density
    per BS held constant at 5 UEs/BS.
    Error bars show 95\% bootstrap confidence intervals.}
    \label{fig:ratebw}
    \end{figure}

    \begin{figure}[!htb]
    \centering
    \includegraphics[width=2.6in]{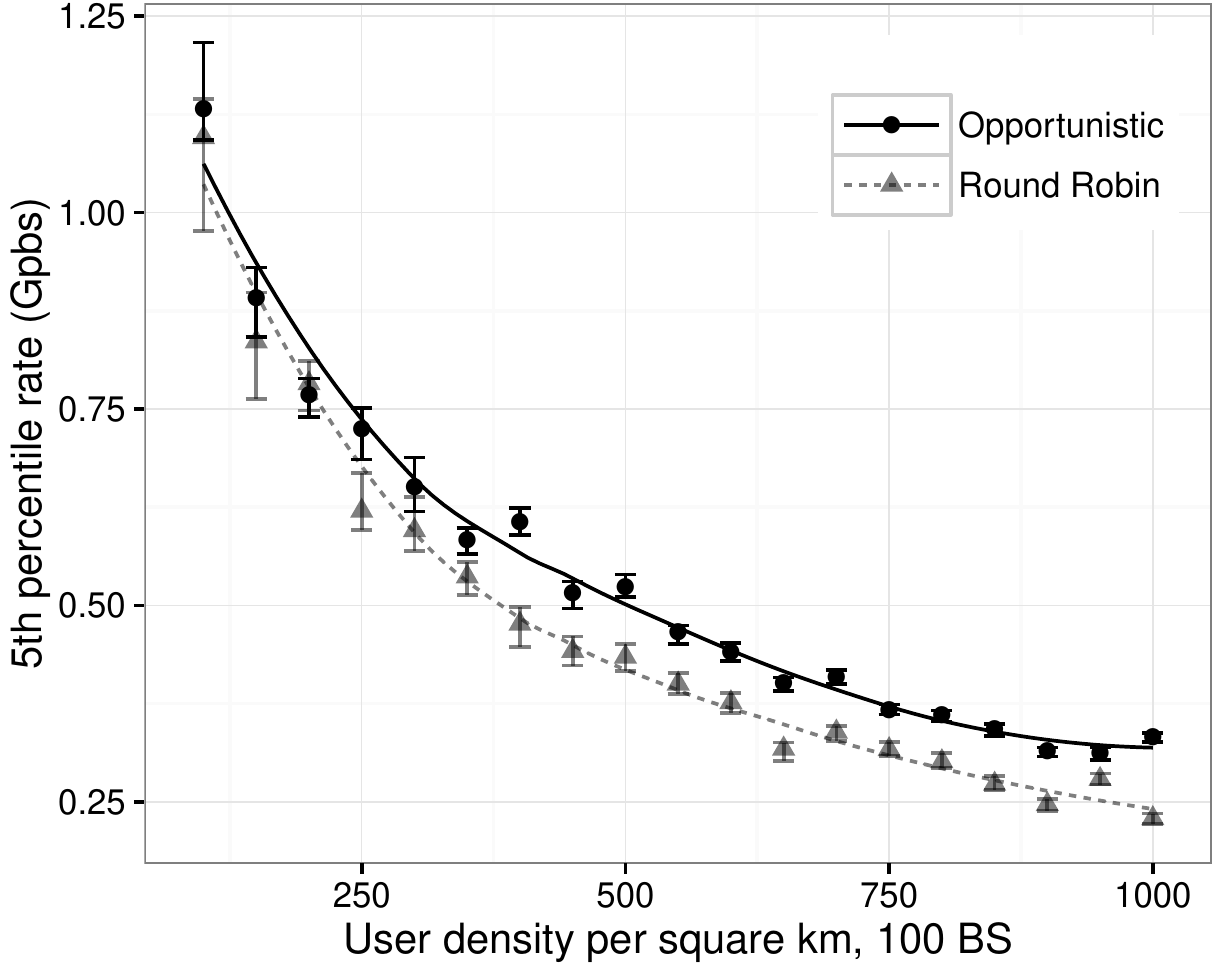}
    \caption[]{Effect of increasing UE density per BS
    as bandwidth is held constant at 1 GHz and BS density is held
    constant at 100 BSs per square kilometer.
    Error bars show 95\% bootstrap confidence intervals.}
    \label{fig:rateuser}
    \end{figure}

\else
    \begin{figure}[!htb]
        \centering
        \begin{minipage}{.46\textwidth}
    \includegraphics[width=2.8in]{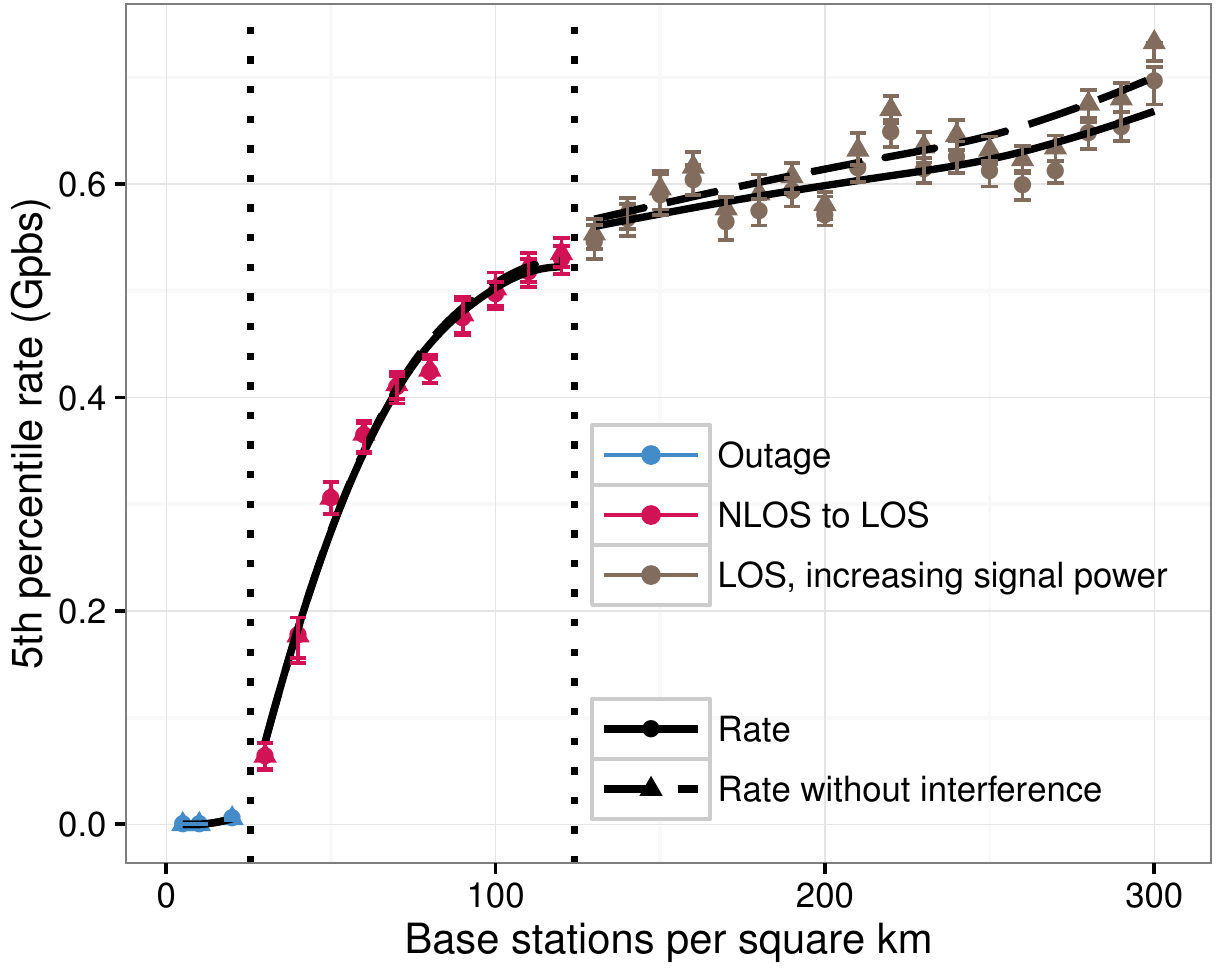}
    \caption[]{Effect of increasing BS density on fifth
    percentile rate as bandwidth is held constant at 1 GHz and UE
    density per BS is held constant at 5 UEs/BS.
    Error bars show 95\% bootstrap confidence intervals.}
    \label{fig:ratebs}
        \end{minipage}%
        \hfill
        \begin{minipage}{0.46\textwidth}
    \includegraphics[width=2.8in]{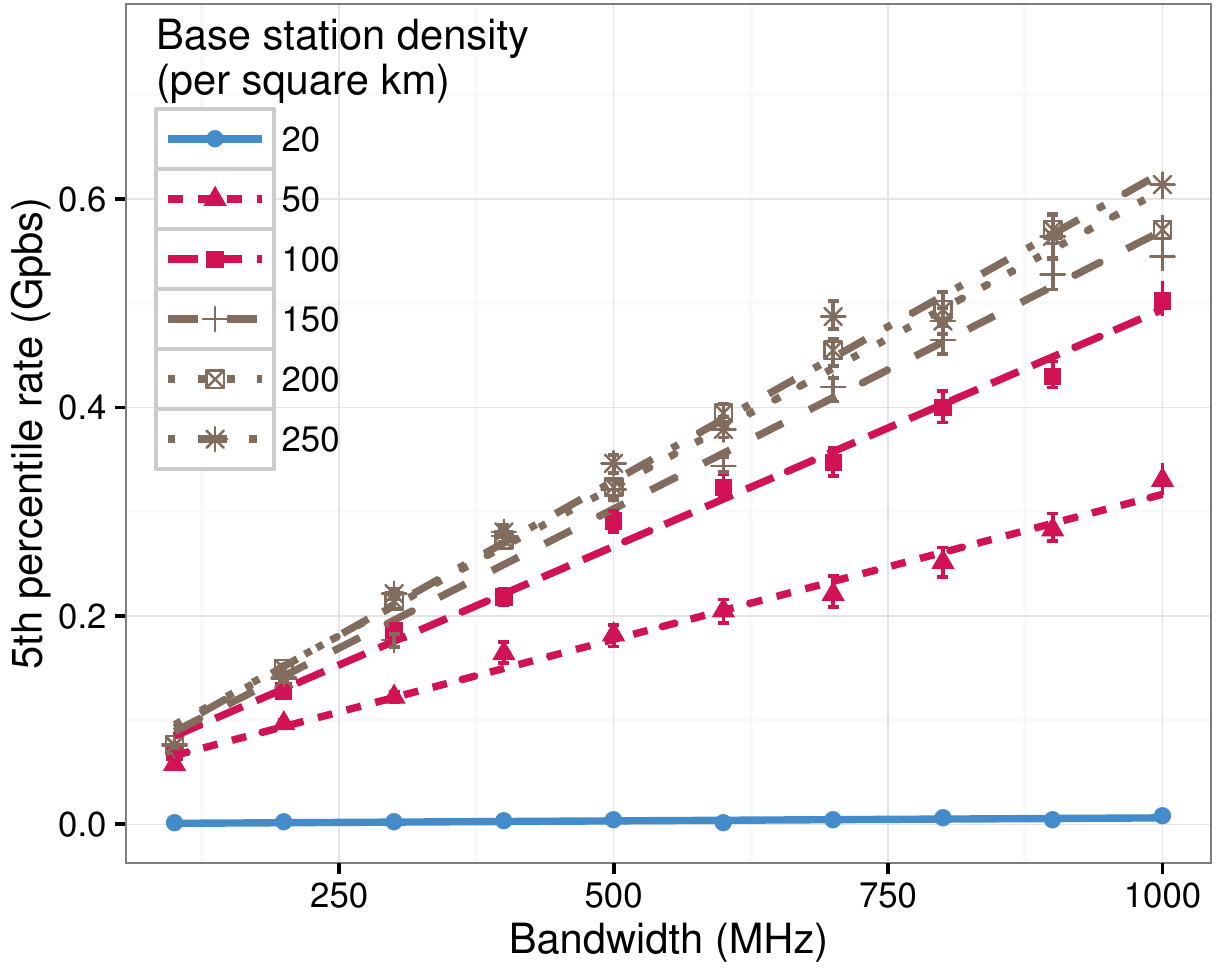}
    \caption[]{Effect of increasing bandwidth on fifth percentile
    rate for different BS densities, with UE density
    per BS held constant at 5 UEs/BS.
    Error bars show 95\% bootstrap confidence intervals.}
    \label{fig:ratebw}
        \end{minipage}
    \end{figure}

    \begin{figure}[h]
    \centering
    \includegraphics[width=2.8in]{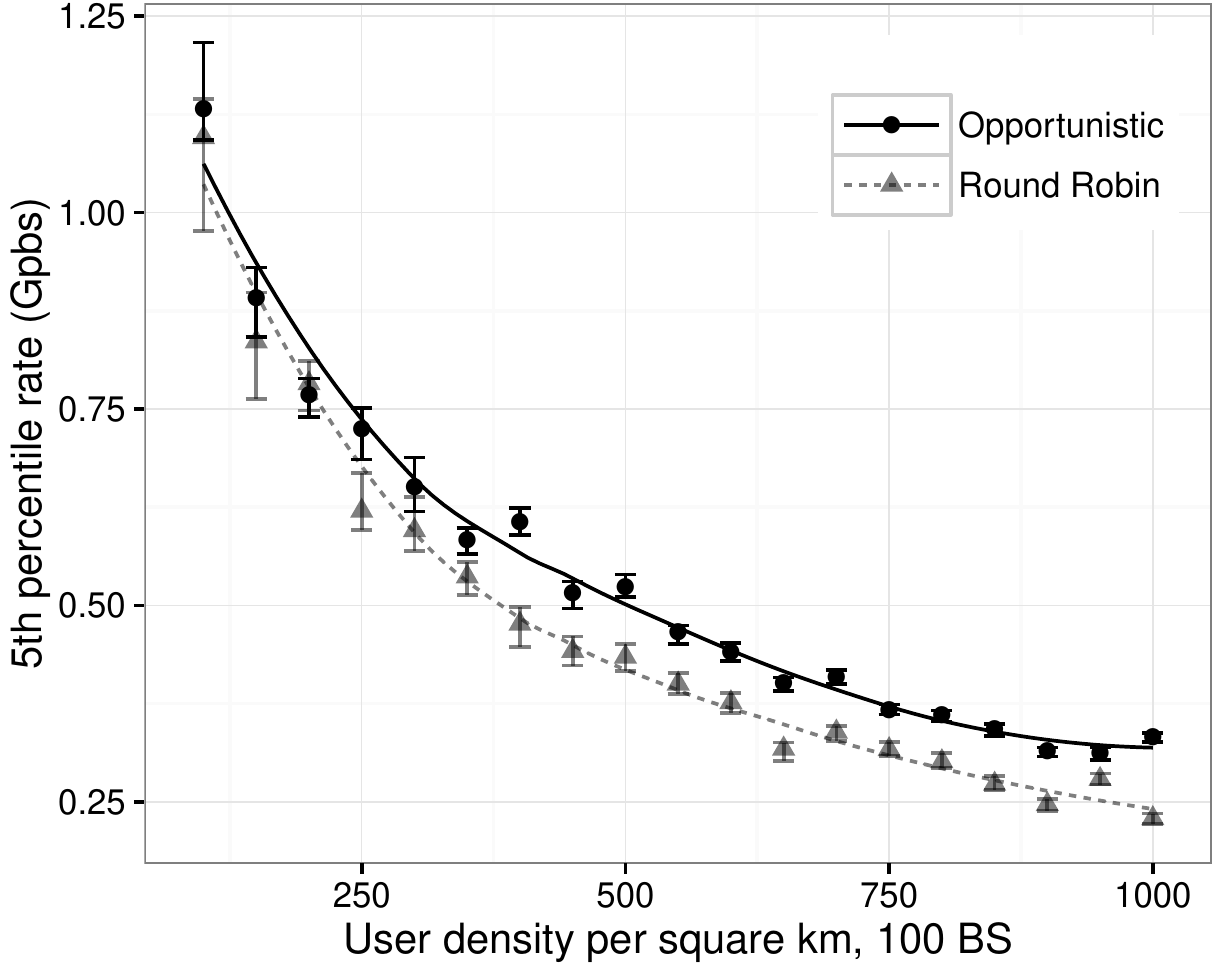}
    \caption[]{Effect of increasing UE density per BS
    as bandwidth is held constant at 1 GHz and BS density is held
    constant at 100 BSs per square kilometer.
    Error bars show 95\% bootstrap confidence intervals.}
    \label{fig:rateuser}
    \end{figure}

\fi
\makeatother

From Fig.~\ref{fig:ratebs},
Fig.~\ref{fig:ratebw}, and Fig.~\ref{fig:rateuser}, 
we find that the fifth percentile data rate of UEs 
is affected by network size as follows:
\begin{enumerate}
\item \textbf{BS density}: (Fig.~\ref{fig:ratebs})
As BS density increases, UEs transition between
three regions, marked by the vertical lines in Fig.~\ref{fig:ratebs}.
In the outage region on the left, the deployment of mmWave BSs
is very sparse, and UEs are likely to be outside the coverage area.
In the middle region, increasing BS density improves the probability
of having a LOS link, and so the rate grows very quickly with BS
density until a point at which virtually all links are LOS. In the third region,
there is a smaller marginal benefit associated with higher BS density
due to increasing SNR. When the deployment of
BSs is extremely dense, there may be a
small negative interference effect that partially mitigates the
positive effect of increasing BS
density, consistent with~\cite{european-sharing}.
\item \textbf{Bandwidth}: (Fig.~\ref{fig:ratebw})
For UEs with a moderate or high SNR, link 
capacity scales linearly with bandwidth,
so these will benefit from subscribing to a large service
provider with more mmWave spectrum holdings.
However, UEs with a very weak signal power (e.g., in a network
where the density of BSs is very low)
are power-limited and do not benefit much from increased bandwidth.
\item \textbf{UE density}: (Fig.~\ref{fig:rateuser})
Given a fixed number of BSs, increasing the
total number of UEs can overburden the network and
create a negative congestion externality. Specifically,
when there are $N$ UEs in the cell,
each UE is allocated approximately $\frac{1}{N}$th
of frequency-time resources. However, because the scheduling of UEs
is partially opportunistic, a UE is more likely
to be scheduled in high-SINR time periods, and this
effect increases with the number of UEs in the cell
as the ``competition'' to be scheduled becomes more intense.
Thus with an increasing number of UEs in a cell, an
individual UE is scheduled in fewer time slots,
but has a higher average data rate in the time slots
in which it \emph{is} scheduled.
\end{enumerate}

In general, the benefit to a UE of increasing any of the 
three elements of network size discussed above 
depends on the UE's signal quality.
In sparse deployments, UEs are outside the coverage area or at its edge,
and derive little or no benefit from increasing bandwidth alone.
Under moderate SNR conditions, UEs
benefit both from increasing BS density
and increasing bandwidth.
In a very dense network, virtually all UEs have a LOS link and high SNR,
and have little to gain from increasing BS density, but 
benefit from increasing bandwidth.

\section{Demand for mmWave services, with and without open resources}
\label{sec:demand}

\begin{figure}[t]
\centering
\includegraphics[width=2.8in]{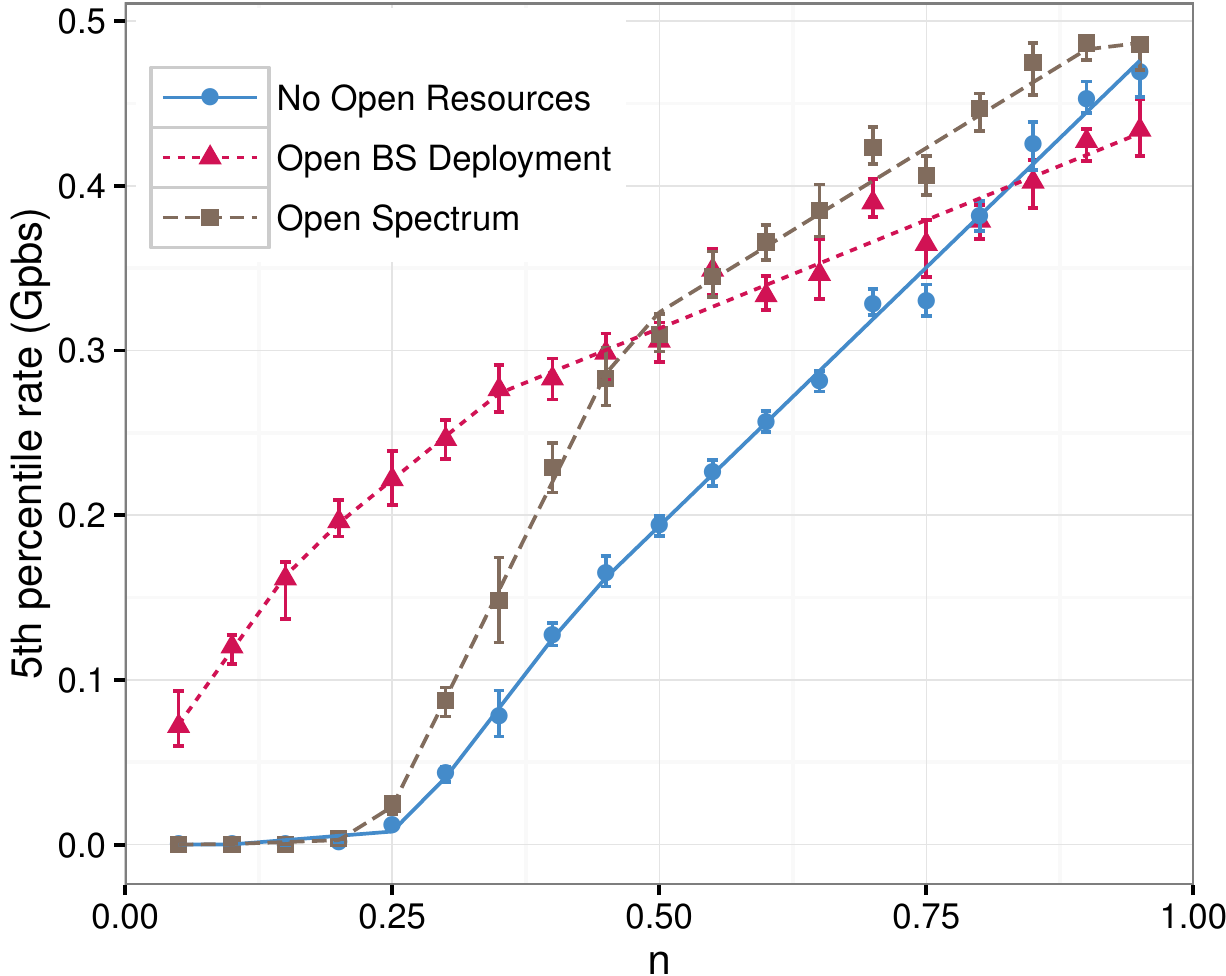}
\caption[]{Effect of increasing network size on fifth percentile UE rate.
Error bars show 95\% bootstrap confidence intervals.}
\label{fig:networksize}
\end{figure}

Having quantified the technical effects on fifth percentile rate of
increasing mmWave network size in Section~\ref{sec:network},
we focus on how demand for wireless service, price an NSP can charge,
and NSP's revenue,
depend on network size.
We are especially interested in the evolution of demand
at small network sizes, when an NSP first begins
to offer mmWave services, and whether the network 
will reach critical mass.
To address this, we model the willingness
to pay of consumers as a function of $n$ (using $h(n)$),
then construct a curve of fulfilled expectations
demand $p(n;n)$ that shows how demand, price, and revenue
scale with $n$. We assume again that consumers decide  
to subscribe or not based on fifth percentile rates, and 
consider the three scenarios (no open resources, 
open BS deployment, and open spectrum) in the ``No sharing''
row in Table~\ref{tab:networksize}.

Fig.~\ref{fig:networksize} shows the simulated
fifth percentile rate for UEs in a mmWave network, 
from which we derive $h(n)$ empirically.
In the open BS deployment scenario, 100 BSs are available
regardless of $n$; in the open spectrum scenario, bandwidth is 
1~GHz at all values of $n$. Otherwise, bandwidth, 
BS density, and the number of UEs in the network 
scale with $n$ up to 1 GHz, 100 BSs,
and 500 UEs, respectively, at $n=1$.
We note that the behavior of the fifth percentile rate
is roughly piecewise linear in $n$, with breakpoints
when we transition between the SNR regions
of Section~\ref{sec:network}.

\begin{figure*}[t]
\centering
\includegraphics[width=0.9\linewidth]{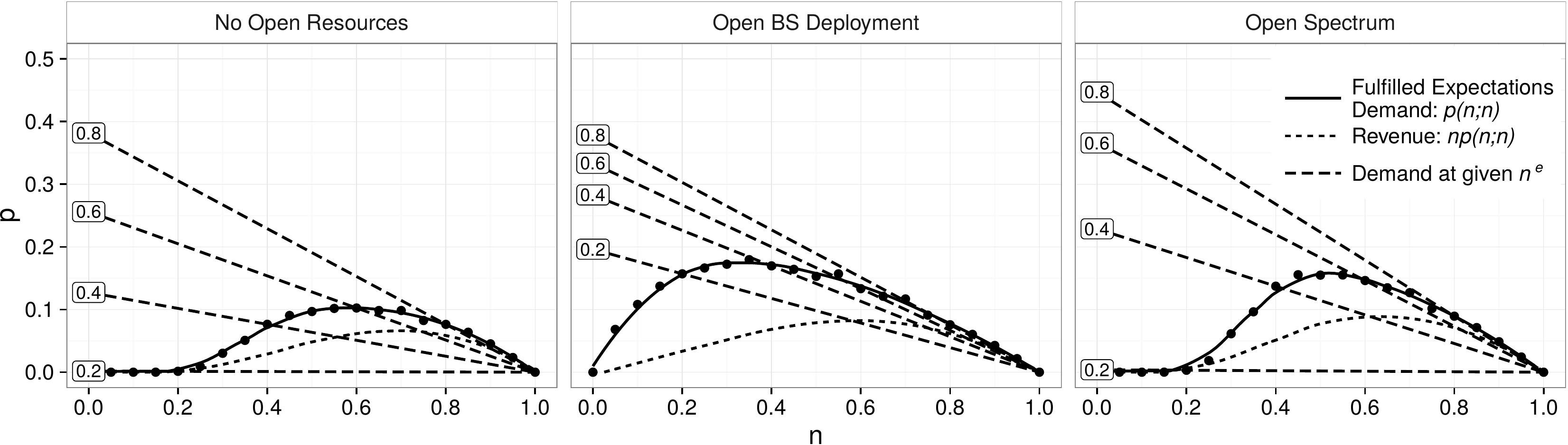}
\caption[]{The fulfilled expectations demand curve $p(n;n)$
and NSP revenue $n p(n;n)$ for a
mmWave network deployment, where the value of the network
externalities function for a network of size $n$, $h(n)$,
is computed by simulation as the fifth percentile rate
of UEs in a network of that size (in Fig.~\ref{fig:networksize}).
We consider three scenarios: one where the size of the network, $n$, 
is proportional to the number of BSs it has
deployed \emph{and} its spectrum resources (left),
one where 100 BSs are open to all networks 
and $n$ is proportional to the network's spectrum resources (middle),
and one where 1 GHz of open spectrum is used by all networks
and $n$ is proportional to the number of BSs 
the network has deployed (right).}
\label{fig:fulfilled}
\end{figure*}

Next, we construct a fulfilled expectations demand
curve for each of the three scenarios, using the values in 
Fig.~\ref{fig:networksize} for the network externalities 
function $h(n)$. 
We assume consumers are heterogeneous
in their willingness to
pay for service, and the parameter $\omega$,
which denotes the consumer type,
is uniformly distributed over $[0,1]$.
A consumer of type $\omega$ subscribing to the network will gain surplus
$u(\omega, n, p)  = \omega h(n) - p$ from a network of size $n$ at price $p$.

The fulfilled expectations demand curve in Fig.~\ref{fig:fulfilled} 
is constructed as described in Section~\ref{sec:economic}.
We also show the revenue of the 
NSP, $n p(n;n)$.
We observe in Fig.~\ref{fig:fulfilled}
the initially upward-sloping fulfilled expectations 
demand curve $p(n;n)$ that is a feature of network goods.

The slope of this curve at small network 
sizes is worthy of extra attention. As described in Section~\ref{sec:economic},
this determines how easily the network 
will reach critical mass in a perfectly competitive market
where competing networks
are homogeneous in every way except for size.
\begin{itemize}
\item When there are no open resources, the slope of $p(n;n)$
is small for $0 \leq n \leq 0.262$, suggesting
that the marginal benefit of increasing network size is
very small  when the network size is small. Under these conditions, 
it is difficult for a network to reach its tipping point.
\item When there is an open BS deployment, 
there is robust demand and strong
marginal network externalities even at very small
network sizes. Assuming a pre-existing BS deployment,
is relatively easy to reach the 
tipping point under these circumstances. 
\item With open spectrum, the positive slope of $p(n;n)$ is 
large for moderate network sizes, but again there is a 
very small positive slope when $n$ is small ($n \leq 0.235$). 
\end{itemize}
We observe that there is a strong marginal network externality
at small network sizes when there is
an open BS deployment,
and NSPs can grow their network by incrementally
adding spectrum holdings and subscribers.
This suggests that based purely on the ability of
a small NSP to generate revenue
(not considering startup costs) and to reach 
its tipping point, an open BS
deployment could ease the barrier to entry for
cellular network providers who are considering extending their
networks to include mmWave service, encouraging
new service offerings. Furthermore, we note that open
BSs help encourage new service offerings more than
open (shared or unlicensed) spectrum would.
However, the open BS scenario relies on a third 
party having invested in BSs, potentially ahead of 
demand if there are no existing mmWave NSPs in the 
market yet and the BSs are only useful for mmWave service.
The open spectrum approach relies only on regulators 
having released the spectrum for unlicensed use in cellular systems.

\section{Resource sharing in a competitive market}
\label{sec:duopoly}

Now we turn our attention to mmWave
NSPs that are already established, 
and are mainly concerned with maximizing their profits
in a competitive market, 
rather than struggling to reach critical mass.
We model the NSPs' decision to share mmWave network resources
or not as a compatibility problem (introduced in Section~\ref{sec:economic}), 
where mmWave NSPs are considered
compatible if their subscribers can connect to
any of the set of NSPs' BSs, and use
a bandwidth equal to their pooled spectrum holdings. 
We previously addressed the
technical benefits of this in Section~\ref{sec:technical}.
We consider all of the combinations of ``shared''
and ``open'' resources in Table~\ref{tab:networksize}.

In many locations in the United States and around the world,
the market for cellular service is effectively a
duopoly. We therefore consider a market
with two vertically differentiated NSPs, 
and the
three-stage game with complete information
described in~\cite{baake2001vertical}:

\begin{enumerate}
\item NSPs $i\in\{1,2\}$ simultaneously choose inherent quality $q_i$ from the
interval $[0, \hat{q}]$.
\item NSPs $i\in\{1,2\}$ simultaneously set price $p_i$.
\item Each consumer chooses to subscribe to one of the NSPs $i\in\{1,2\}$
or to neither.
\end{enumerate}

The quality $q_i$ is the \emph{inherent} quality (with maximum feasible value $\hat{q}$).
It refers to aspects of service unrelated to the
size of the mmWave network, such as the quality of voice calls,
the quality of the legacy data network,
customer service, and the availability of desirable handsets.
These measures of quality are increasingly 
important once the service is well established in the market,
and has a sufficiently large network that network size 
is no longer the single main criterion by which consumers
decide which NSP to subscribe to.

An NSP's marginal costs are increasing in $q_i$,
with cost function
$c(q_i, n_i) = q_i n_i$, 
and so each NSP $i \in \{1,2\}$ seeks to maximize its profits
\begin{equation}
\pi_i(q_i, n_i, p_i) = n_i p_i - q_i n_i
\end{equation}

Consumers evaluate competing services in terms of the difference in
their inherent qualities as well as their network externalities.
We have heterogeneous consumers parameterized by $\omega$, 
with $\omega$ distributed uniformly from $[0, \hat{\omega}]$.
The surplus of a consumer of type $\omega$ is given by
\makeatletter
\if@twocolumn%
    \begin{equation}
     u(\omega, q_i, \tilde{n}_i, p_i ) =
      \begin{cases}
       \omega q_i + \mu q_i \tilde{n}_i - p_i & \text{if subscribes to } i  \\
       0       & \text{if no subscription}
      \end{cases}
    \end{equation}

\else
    \begin{equation}
     u(\omega, q_i, \tilde{n}_i, p_i ) =
      \begin{cases}
       \omega q_i + \mu q_i \tilde{n}_i - p_i & \text{if the consumer subscribes to network } i  \\
       0       & \text{if the consumer does not subscribe to a network}
      \end{cases}
    \end{equation}
\fi
\makeatother
\noindent with $i \in \{1,2\}$ and $0 \leq \mu < \text{min}[1, \hat{\omega}/2]$, 
where $\mu$ is explained in the next paragraph.
If the NSPs share their mmWave network resources,
then $\tilde{n}_i = \sum_{i \in \{1,2\}} n_{i}$, otherwise $\tilde{n}_i = n_i$.

The fifth percentile rate in a mmWave network is piecewise linear 
in the network size (Fig.~\ref{fig:networksize}). 
Here we consider only moderate- to large-sized
networks, where the curves in 
Fig.~\ref{fig:networksize} are linear. 
Thus the network externalities 
function $h(\tilde{n}_i)$ is linear in $\tilde{n}_i$.
The scaling factor $\mu$ determines the intensity of the network
externality, i.e.,  $h(\tilde{n}_i) = \mu \tilde{n}_i$, 
and is empirically derived from slopes of the lines
in Fig.~\ref{fig:networksize}. 
We consider three scenarios, each with established
NSPs:
\begin{itemize}
\item \textbf{No open resources}: In this scenario we 
use $\mu=0.7$, corresponding to the slope of the ``no open resources''
line in Figure~\ref{fig:networksize} for $n \geq 0.25$.
\item \textbf{Open BS deployment}: There is an 
open BS deployment serving all NSPs. We use $\mu=0.25$, 
corresponsing to the slope of the ``open BS deployment''
line in Fig.~\ref{fig:networksize} for $n \geq 0.35$.
\item \textbf{Open spectrum}: Spectrum is unlicensed and used
by all NSPs. We use $\mu=0.4$, 
corresponsing to the slope of the ``open spectrum''
line in Fig.~\ref{fig:networksize} for $n \geq 0.45$.
\end{itemize}

By their choice of quality level, the NSPs
segment the market into a low-end group (small-$\omega$ type)
and a high-end group (large-$\omega$ type).
Without loss of generality, we say that NSP 1 chooses a higher
quality than NSP 2, i.e., $q_1 > q_2$, and
subscribers of NSP 1 belong to the
large-$\omega$ group.
We define two marginal consumers: the consumer of type
$\underline{\omega}$ is indifferent between choosing no subscription
and subscribing to NSP 2,
and the consumer of type $\overline{\omega}$ is indifferent between
subscribing to NSP 1 and subscribing to NSP 2.

Then the utility of the marginal consumer of type $\overline{\omega}$
satisfies
\begin{equation}
\overline{\omega} q_1 + \mu q_1 \tilde{n}_1 - p_1  = \overline{\omega} q_2 + \mu q_2 \tilde{n}_2 - p_2
\label{omegaover}
\end{equation}
and the utility of the marginal consumer of type $\underline{\omega}$
satisfies
\begin{equation}
\underline{\omega} q_2 + \mu q_2 \tilde{n}_2 - p_2  = 0
\label{omegaunder}
\end{equation}

Also, the marginal consumer of type $\overline{\omega}$
defines the market share of the high-end service
\begin{equation}
n_1 = \frac{\hat{\omega}-\overline{\omega}}{\hat{\omega}}
\label{n1share}
\end{equation}
and the marginal consumers together define
the market share of the low-end service
\begin{equation}
n_2 = \frac{\overline{\omega}-\underline{\omega}}{\hat{\omega}}
\label{n2share}
\end{equation}

We can solve (\ref{omegaover}), (\ref{omegaunder}), (\ref{n1share}), and (\ref{n2share})
for $n_1$, $n_2$, $\overline{\omega}$, and $\underline{\omega}$,
and thus determine the decisions of the consumers and the market
share of each NSP given $p_i, q_i, i \in \{1,2\}$.

Given $p_i$ and $q_i$~$i\in\{1,2\}$, it is 
shown in \cite{baake2001vertical} that
if the ratio of quality levels satisfies
\begin{equation}
\frac{q_1}{q_2} > \Bigg( \frac{\hat{\omega}^2}{(\hat{\omega}-\mu)(\hat{\omega}-2\mu)}\Bigg)
\label{highprice}
\end{equation}
then there is a unique Nash equilibrium in which both
NSPs set prices higher than their marginal costs.
Furthermore, if the solution to (\ref{omegaover}),
(\ref{omegaunder}), (\ref{n1share}), and (\ref{n2share}) satisfies
\begin{equation}
0 < \underline{\omega} < \overline{\omega} < \hat{\omega}
\label{marketshare}
\end{equation}
then both NSPs have market share greater than zero.
When both (\ref{highprice}) and (\ref{marketshare}) hold, then there is a unique
Nash equilibrium in which both NSPs earn non-zero profit.
We restrict our attention to these circumstances, since these
are of primary interest to us.

If (\ref{highprice}) and (\ref{marketshare}) hold and
the NSPs do not share resources,
then according to \cite{baake2001vertical} their equilibrium
prices $p_{1,NS}^*, p_{2,NS}^*$ are as follows:
\begin{equation}
p_{1,NS}^* = q_1 \Bigg[1 + \frac{(\hat{\omega}-1)[2q_1(\hat{\omega}-\mu)^2 - q_2 \hat{\omega}(2\hat{\omega}-\mu)]}{4q_1(\hat{\omega}-\mu)^2 - q_2\hat{\omega}^2} \Bigg] > q_1
\end{equation}
\begin{equation}
p_{2,NS}^* = q_2 \Bigg[1 + \frac{(\hat{\omega}-1)[q_1(\hat{\omega}-\mu)(\hat{\omega}-2\mu) - q_2\hat{\omega}^2]}{4q_1(\hat{\omega}-\mu)^2 - q_2\hat{\omega}^2} \Bigg] > q_2
\end{equation}

\noindent and their equilibrium
quality levels $q_{1,NS}^*, q_{2,NS}^*$ are:
\begin{equation}
q_{1,NS}^* = \hat{q}
\end{equation}

\makeatletter
\if@twocolumn%
    \begin{equation}
    \resizebox{0.42\textwidth}{!}{
    $q_{2,NS}^* = \frac{\hat{q}(\hat{\omega}-\mu)^2 \Big[11\hat{\omega}-10\mu - \sqrt{3(3\hat{\omega}^2 + 28\hat{\omega}\mu - 20\mu^2)}\Big]}{2\hat{\omega}^2(7\hat{\omega}-5\mu)}  < \hat{q}$
    }
    \end{equation}
\else
    \begin{equation}
    q_{2,NS}^* = \frac{\hat{q}(\hat{\omega}-\mu)^2 \Big[11\hat{\omega}-10\mu - \sqrt{3(3\hat{\omega}^2 + 28\hat{\omega}\mu - 20\mu^2)}\Big]}{2\hat{\omega}^2(7\hat{\omega}-5\mu)}  < \hat{q}
    \end{equation}
\fi
\makeatother

The profits of the high-end NSP always increase with $q_1$, 
so it will use $\hat{q}$. The low-end NSP sets $q_2$ to balance
two competing effects: at high values of $q_2$ the low-end NSP 
has a greater market share, but is also more similar to $q_1$, which 
increases price competition and drives the price of service down.

Note that reducing the parameter $\hat{q}$ increases price competition,
since the difference in quality levels between NSPs
will be small and so consumers' decisions will be more sensitive to price.
Similarly, reducing $\hat{\omega}$
increases price competition, since this decreases the dispersion
of consumers' willingness to pay and the market is less segmented.

If the NSPs share resources,
then per \cite{baake2001vertical} their equilibrium
prices $p_{1, S}^{*}, p_{2, S}^{*}$ are:
\begin{equation}
p_{1, S}^{*} = q_1 \Bigg[ 1 + \frac{2\hat{\omega}(\hat{\omega}-1)(q_1-q_2)}{(4\hat{\omega}-3\mu)q_1 - \hat{\omega}q_2} \Bigg] > q_1
\end{equation}
\begin{equation}
p_{2, S}^{*} = q_2 \Bigg[ 1 + \frac{\hat{\omega}(\hat{\omega}-1)(q_1-q_2)}{(4\hat{\omega}-3\mu)q_1 - \hat{\omega}q_2}  \Bigg] > q_2
\end{equation}
\noindent and their equilibrium
quality levels $q_{1,S}^{*}, q_{2,S}^{*}$ are:
\begin{equation}
q_{1,S}^{*} = \hat{q}
\end{equation}
\begin{equation}
q_{2,S}^{*} =  \frac{\hat{q}(4\hat{\omega}-3\mu)}{7\hat{\omega}-6\mu} < \hat{q}
\end{equation}

For the sake of comparison, we are also interested in the
profits of a monopoly NSP. When there is only one NSP,
the marginal consumer is defined by
\begin{equation}
\overline{\omega} q_1 + \mu q_1 \tilde{n}_1 - p_1  = 0
\end{equation}

\noindent and the market share of the NSP is
\begin{equation}
n_1 = \frac{\hat{\omega}-\overline{\omega}}{\hat{\omega}}
\end{equation}
At equilibrium, the monopoly NSP will choose price
\begin{equation}
p_{1,M}^{*} = \frac{q_1(\hat{\omega}-1)}{2}
\end{equation}
and quality level
\begin{equation}
q_{1,M}^{*} = \hat{q}
\end{equation}

\begin{figure*}[ht]
\centering
\includegraphics[width=6.75in]{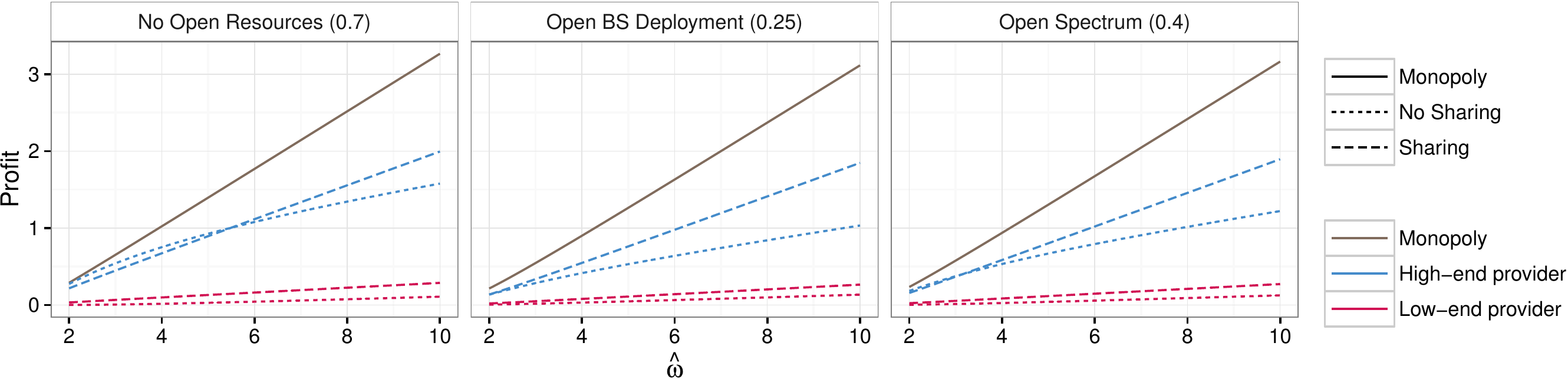}
\caption[]{Profit of each NSP for different cases of the 
intensity of the network effect, $\mu$. The dotted line shows the profits 
of each NSP if they choose not to share their respective 
resources (though they may still use any available ``open'' resources). 
The dashed line shows their profits if they 
share resources. The solid line shows the profits in a monopoly market, 
for comparison. The horizontal axis 
indicates the dispersion of the consumers' valuation of the services. ($\hat{q} = 1.5$)}
\label{fig:duopolyprofit}
\end{figure*}

Fig.~\ref{fig:duopolyprofit} shows the profits of each NSP 
in various circumstances, as $\hat{\omega}$ (and dispersion
of consumers' willingness to pay) increases.
First, we note that the low-end NSP always prefers to share 
resources. Since it captures the low end of the market (the consumers
who are less willing to pay for inherent quality), the 
network effects are especially important to this NSP.
In the duopoly market, the high-end NSP paradoxically prefers to share 
resources when the intensity of the network effect is small
(as when there are open resources), 
because its competitor gains
less of an advantage from the larger (shared)
network size. (This is consistent with the results described
in~\cite{economides1998equilibrium}). When the intensity of the network 
effect is large (as it is when there are no open resources) then 
the high-end NSP prefers resource sharing only
when the market is highly segmented and there is little price competition
(i.e., for large $\hat{\omega}$).

\begin{figure*}[ht]
\centering
\includegraphics[width=6.75in]{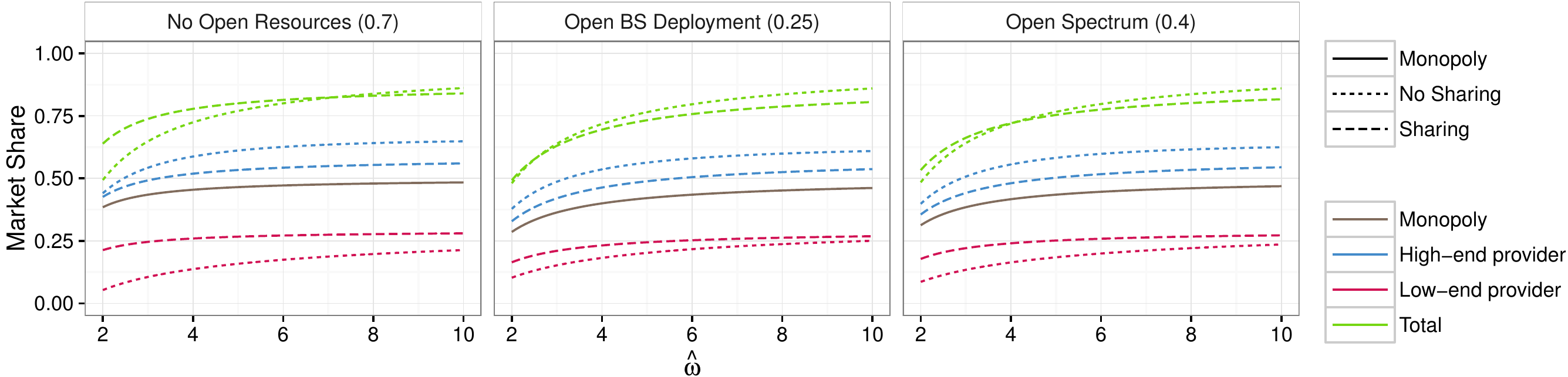}
\caption[]{Market share of each NSP and total market coverage 
for different cases of the 
intensity of the network effect, $\mu$. The dotted line shows the market share 
of each service provider and their total market coverage 
if they choose not to share their respective 
resources (though they may still use any available ``open'' resources). 
The dashed line shows their market shares and total market coverage if they  
share resources. The solid line shows the market share of a single
firm in a monopoly market, for comparison. The horizontal axis 
indicates the dispersion of the consumers' valuation of the services. ($\hat{q} = 1.5$)}
\label{fig:duopolymarket}
\end{figure*}

Fig.~\ref{fig:duopolymarket} shows the market 
share of each NSP and total market coverage (i.e., share of 
consumers who subscribe to either NSP) under
the same set of circumstances. When $\hat{\omega}$ (and 
the dispersion of consumers' willingness to pay) is small, 
sharing offers the best overall market coverage.
When $\hat{\omega}$ is large, the best market coverage is achieved
by not sharing resources. The value of $\hat{\omega}$ at 
which the benefit of market segmentation begins to dominate 
the benefit of network effects is greater when the intensity 
of the network effect is high (large $\mu$).

\section{Conclusions}
\label{sec:conclusion}

In this paper, we have connected economic models of the strategic 
decision making of cellular network service providers 
and subscribers, to detailed simulations of mmWave 
networks, with and without resource sharing.
While we have confirmed the benefits of resource sharing 
from a purely technical view (without considering 
the effect on demand), with the economic analysis we 
have illustrated that resource sharing 
is not always the preferred strategy of service
providers, and some kinds of resource sharing may be preferred over others.
We have shown that ``open'' deployments of neutral small cells 
make it easier for networks to reach critical mass, encouraging market
entry more than ``open'' spectrum would. 
Furthermore, we have shown that the leading service provider
in a duopoly market prefers to share resources only when sharing gains 
are small or the market is highly segmented.
Our technical simulations of asymmetric service providers have hinted at this,
with greater gains for the smaller service provider than the market leader.
However with a purely technical approach, one would conclude that 
resource sharing is always beneficial (albeit less beneficial for the 
market leader), while the economic analysis with consideration of
price and demand in addition to technical gains 
has suggested a different conclusion.

We briefly discuss here some assumptions of our approach.
Our results are predicated on an assumed indirect
network effect benefitting consumers subscribing to
a large service provider.
That is, we assume that the
resources held by a service provider in a given market
scale together with the number of subscribers it serves.
Practically, building out physical infrastructure and licensing
spectrum requires a tremendous capital investment. A service
provider is unlikely to build out a very large network, at great
cost, when it has few subscribers and so a limited revenue stream.
For this reason, we consider it justified to tie the level of
investment in the network - and thus, the size of the network resources -
to the number of subscribers.
Another assumption is that consumers are homogeneous in their preference
for one firm or the other, given their overall valuation of network
service, i.e., that consumers with the same $\omega$
will make the same choice between service providers, given their
price, network size, and inherent quality. Actually, consumers
and are not identical in their
valuations of competing services. However, despite this common 
simplifying assumption, the general economic framework we have 
applied in this paper has been empirically validated 
in a variety of other industries with network effects.

The work in this paper suggests several interesting avenues for further research.
We would like to extend
this model to include the investment costs associated with deploying a
new mmWave network, which we expect to be substantially different 
from traditional cellular networks given the unique physical characteristics
of the mmWave bands.
We would also like to investigate scheduling strategies
that divide shared resources among mmWave service providers 
in ways that increase the benefit of resource sharing for 
both service providers and consumers.

\ifCLASSOPTIONcaptionsoff
  \newpage
\fi

\bibliographystyle{IEEEtran}
\bibliography{jsac}

\end{document}